\documentclass[
aps,
prl,
twocolumn,
showpacs,
amsmath,
amssymb,
superscriptaddress,
longbibliography
]{revtex4-2}
\usepackage[utf8]{inputenc}
\usepackage{graphicx,graphics}
\usepackage{subfigure} 
\usepackage{mathrsfs} 
\usepackage{hyperref}
\usepackage{tikz}
\usetikzlibrary{bending,hobby}
\usepackage[compat=1.1.0]{tikz-feynman}
\usetikzlibrary{arrows.meta}
\tikzfeynmanset{
  warn luatex=false,
  my arrow/.style 2 args={
    decoration={
      markings,
      mark=at position #1 with {
        \node[
        transform shape,
        fill=#2,
        inner sep=1.4pt,
        draw=none,
        dart
        ] { };
      },
    },
    postaction={
      decorate=true,
    },
  },
  fs/.style={draw,line width=1pt,color=black!100,
    my arrow={0.5}{black!100}},
  ff/.style={draw,line width=0.5pt,color=black!70,
    my arrow={0.5}{black!50},dash=on 1pt off 1pt phase 0},
  my boson/.style 2 args={
    draw=none,
    decoration={name=none},
    postaction={
      draw,
      line width=#1,
      color=#2,
      decoration={
        snake,
        amplitude = 1,
        segment length = 5,
        pre=curveto,
        pre length=1,
        post=curveto,
        post length=0,
      },
      decorate=true,
    },
  },
  bs/.style={my boson={1pt}{black!100},my arrow={0.5}{black!100}},
  bf/.style={my boson={0.5pt}{black!70},my arrow={0.5}{black!50},
    dash=on 0.5pt off 0.5pt phase 0,line width=0.5pt},
  double arrows/.style n args={4}{
    decoration={
      markings,
      mark=at position 0.5 with {
        \arrow[xshift=7.0pt]{
          Stealth[#1,length=#3,sep=#4]
          Stealth[#2,length=#3]
        }
      },
    },
    postaction={
      decorate=true,
    },
  },
  normal GF/.style = {
    draw,
    line width=0.5pt,
    color=black!70,
    dash=on 1pt off 1pt phase 0,
    double arrows = {}{}{6pt}{2pt}
  },
  anomalous GFin/.style = {
    draw,
    line width=0.5pt,
    color=black!70,
    dash=on 1pt off 1pt phase 0,
    double arrows = {}{reversed}{6pt}{1.5pt}
  },
  anomalous GFout/.style = {
    draw,
    line width=0.5pt,
    color=black!70,
    dash=on 1pt off 1pt phase 0,
    double arrows = {reversed}{}{6pt}{2pt}
  },
}

\newcommand{\DefLegShort}[5]{
  \hspace{0em}\raisebox{3pt}{
    \begin{tikzpicture}[baseline=(v0)]
      \begin{feynman}[small]
        \vertex [dot] (v0) at (0,0) {};
        \vertex [#3 = 0.8cm of v0] (v) {};
        \diagram{#1 -- [#2,#5] #4};
      \end{feynman}
    \end{tikzpicture}
  } \hspace{0em}
}

\newcommand{\DefLegLong}[4]{
  \hspace{-0.3em}\raisebox{3pt}{
    \begin{tikzpicture}[baseline=(v0)]
      \begin{feynman}[small]
        \vertex [dot] (v0) at (0,0) {};
        \vertex [right = 1.3cm of v0] (v) {};
        \diagram{#1 -- [#2,#3] #4};
      \end{feynman}
    \end{tikzpicture}
  } \hspace{-0.7em}
}

\newcommand{\DefImpBoseInt}[4]{
  \hspace{-0.5em}\raisebox{-0pt}{
    \begin{tikzpicture}[baseline=(v)]
      \begin{feynman}[small]
        \vertex [dot] (v) at (0,0) {};
        \vertex [left = 0.5 of v] (l) {};
        \vertex [right = 0.5 of v] (r) {};
        \vertex [above = 0.8 of v,crossed dot] (p) {};
        \diagram{
          (r) -- [#1,edge label'=#2] (v),
          (v) -- [#3,edge label'=#4] (l),
          (p) -- [my boson={1pt}{black!100}] (v),
        };
      \end{feynman}
    \end{tikzpicture}
  }\hspace{-0.5em}
}

\newcommand{\LI}[2]{
  \hspace{-0.5em}\raisebox{-8pt}{
    \begin{tikzpicture}[baseline=(v)]
      \begin{feynman}[small]
        \vertex [dot] (v) at (1,0) {};
        \vertex [left = of v] (l) {};
        \vertex [right = of v] (r) {};
        \vertex [above = 0.8 of v,crossed dot] (p) {};
        \diagram{
          (r) -- [#1,edge label'=\(p_{2}\)] (v),
          (v) -- [#2,edge label'=\(p_{1}\)] (l),
          (p) -- [my boson={1pt}{black!100}] (v),
        };
      \end{feynman}
    \end{tikzpicture}
  }\hspace{-0.5em}
}

\newcommand{\SsqContribOne}[7]{
  \hspace{-0.5em}\raisebox{-8pt}{
    \begin{tikzpicture}[baseline=(v1)]
      \begin{feynman}[small]
        \vertex [dot] (v1) at (0.5,0) {};
        \vertex [dot, right = 1 of v1] (v2) {};
        \vertex [left = 0.7 of v1] (l) {};
        \vertex [right = 0.7 of v2] (r) {};
        \vertex [above = 0.8 of v1,crossed dot] (p1) {};
        \vertex [above = 0.8 of v2,crossed dot] (p2) {};
        \diagram{
          #1 -- [#2] #3,
          (v2) -- [#4] (v1),
          #5 -- [#6] #7,
          (p1) -- [my boson={1pt}{black!100}] (v1),
          (p2) -- [my boson={1pt}{black!100}] (v2),
        };
      \end{feynman}
    \end{tikzpicture}
  }\hspace{-0.5em}
}

\newcommand{\DefTwoPtGreenFun}[4]{
  \hspace{0em}\raisebox{3pt}{
    \begin{tikzpicture}[baseline=(v1)]
      \begin{feynman}[small]
        \vertex [dot] (v1) at (0.5,0) {};
        \vertex [dot, right = 0.8 of v1] (v2) {};
        \diagram{
          #1 -- [#2,#4] #3,
        };
      \end{feynman}
    \end{tikzpicture}
  }\hspace{-0.3em}
}

\newcommand{\FDZeroLegOneLoop}{
  \hspace{-0.6em}\raisebox{3pt}{
    \begin{tikzpicture}[baseline=(v0)]
      \begin{feynman}[small]
        \vertex [dot] (v0) at (0,0) {};
        \vertex [right = 0.7 of v0, crossed dot] (v) {};
        \vertex at (-0.5,0.5) (p1) {\(p_{1}\)};
        \vertex at (-0.5,-0.5) (p2) {\(p_{2}\)};
        \diagram{
          (v0) -- [my boson={1pt}{black!100}] (v),
          v0 -- [normal GF, out=135,in=-135,min distance=1.6cm] v0,
        };
      \end{feynman}
    \end{tikzpicture}
  } \hspace{0em}
}

\hypersetup{
colorlinks=true,     
linkcolor=blue,     
citecolor=blue,     
urlcolor=blue,     
}
\newcommand{\ud}{\mathrm{d}}
\newcommand{\rRe}{\mathrm{Re}}
\newcommand{\rIm}{\mathrm{Im}}
\newcommand{\sgn}{\mathrm{sgn}}
\newcommand{\br}{\mathrm{\mathbf{r}}}
\newcommand{\bR}{\mathrm{\mathbf{R}}}
\newcommand{\bk}{\mathrm{\mathbf{k}}}
\newcommand{\bp}{\mathrm{\mathbf{p}}}
\newcommand{\bq}{\mathrm{\mathbf{q}}}
\newcommand{\aphi}{\phi^{\ast}}
\newcommand{\s}{\mathrm{s}}
\newcommand{\phis}{\phi^{(\s)}}
\newcommand{\aphis}{\phi^{(\s)\ast}}
\newcommand{\f}{\mathrm{f}}
\newcommand{\phif}{\phi^{(\f)}}
\newcommand{\aphif}{\phi^{(\f)\ast}}

\newcommand{\cL}{\mathcal{L}}
\newcommand{\cD}{\mathcal{D}}

\newcommand\sect[1]{\paragraph{#1.---\hspace{-1em}}}

\begin{document}

\title{Asymptotic Freedom of Two Heavy Impurities in a Bose-Einstein
  Condensate}

\author{Dong-Chen Zheng}
\affiliation{Fujian Provincial Key Laboratory for Quantum Manipulation and New Energy Materials, College of Physics and Energy, Fujian Normal University, Fuzhou 350117, China}
\affiliation{Fujian Provincial Collaborative Innovation Center for Advanced High-Field Superconducting Materials and Engineering, Fuzhou, 350117, China}
\author{Lin Wen}
\email{wlqx@cqnu.edu.cn}
\affiliation{College of Physics and Electronic Engineering, Chongqing Normal University, Chongqing 401331, China}
\author{Renyuan Liao}
\email{ryliao@fjnu.edu.cn}
\affiliation{Fujian Provincial Key Laboratory for Quantum Manipulation and New Energy Materials, College of Physics and Energy, Fujian Normal University, Fuzhou 350117, China}
\affiliation{Fujian Provincial Collaborative Innovation Center for Advanced High-Field Superconducting Materials and Engineering, Fuzhou, 350117, China}
\date{\today}

\begin{abstract}
We consider two heavy impurities immersed in a Bose-Einstein
condensate, and calculate the self-energy using the Wilsonian
renormalization.
The polaron energy, quasiparticle residue and damping
rate are extracted from the self-energy.
We demonstrate that various effective potentials emerge from the
polaron energy under the specific conditions.
In the limit of large separation between the impurities, the polaron
spectrum converges to the results for a single impurity, exhibiting an
attractive-repulsive crossover across the Feshbach resonance.
The boundary of this crossover is identified through the analysis of
the damping rate.
We highlight that repulsive-dominant polarons can exist as long as
the impurities are sufficiently close, even when the impurity-boson
interactions are attractive.
Additionally, we observe that the two impurities become asymptotically
free in the repulsive polaron regime.
These results are verifiable and offer a fresh perspective on the
interaction dynamics between two polarons.

\end{abstract}

\maketitle


\sect{Introduction}

Investigating particle interactions is a cornerstone of modern physics
as these interactions fundamentally dictate the behavior of matter and
energy.
A crucial concept within this framework is the mediating interactions.
For instance, mesons mediate nuclear forces between
baryons~\cite{10.1143/PTPS.1.1}, photons mediate electromagnetic
interactions between charged
particles~\cite{doi:10.1098/rspa.1927.0039},
and gluons mediate strong interactions between
quarks~\cite{PhysRev.96.191, PhysRevLett.30.1343,
  PhysRevLett.30.1346}.
In condensed matter, phonon exchange mediates attractive interactions
between electrons, a process vital for
superconductivity~\cite{PhysRev.108.1175}, and electron-phonon
combinations lead to polaron
quasiparticles~\cite{landau1948effective}.
Ultracold atomic systems, utilizing imbalanced quantum gases where a
minority impurity component interacts with a majority environment,
offer a unique platform to study polaron
physics~\cite{RevModPhys.80.885,
  grusdt2024impuritiespolaronsbosonicquantum, Baroni2024NatRevPhys,
  Baroni2024NatPhys}.
Interactions in these systems can be tuned with Feshbach resonances,
and the majority component's quantum statistics determine the polaron
type, leading to Fermi~\cite{PhysRevLett.102.230402,
  PhysRevLett.103.170402, Hannaford2012, Koschorreck2012,
  Kohstall2012, PhysRevX.10.041019, PhysRevLett.125.133401,
  PhysRevA.106.063306} or Bose polarons~\cite{PhysRevLett.117.055301,
  PhysRevLett.117.055302, PhysRevA.99.063607,
  doi:10.1126/science.aax5850, PhysRevResearch.4.043093, Yan2024}.
The Bose polaron, particularly relevant for understanding strongly
correlated electron systems, has attracted much
attention~\cite{PhysRevA.88.053632, PhysRevA.90.013618,
  PhysRevA.92.033612, PhysRevA.103.033317, PhysRevA.104.023317,
  PhysRevA.104.052218, PhysRevLett.127.205301, PhysRevA.106.033321,
  PhysRevA.109.013325, PhysRevA.110.023310},
with investigations revealing attractive and repulsive branches across
Feshbach resonances~\cite{PhysRevA.88.053632, PhysRevA.90.013618,
  PhysRevA.92.033612}.

The two-impurity problem in superfluids has recently attracted
significant attention in the field of cold atomic
physics~\cite{PhysRevLett.121.080405, PhysRevLett.121.013401,
  PhysRevX.8.031042, doi:10.7566/JPSJ.87.043002, Jager_2022,
  PhysRevA.107.063301, PhysRevA.108.L051301, PhysRevLett.129.233401,
  PhysRevA.110.030101, Panochko_2021, atoms10010019}.
This system provides a valuable platform for studying inter-impurity
forces mediated by collective excitations within the superfluid
medium.
In an ideal Bose-Einstein condensate (BEC), the interaction between
two static impurities is characterized by a ``shifted Newtonian''
attractive potential~\cite{PhysRevA.107.063301}.
In a weakly interacting BEC, the strong impurity-boson interactions
and the short impurity separations lead to the Efimov
potential~\cite{doi:10.7566/JPSJ.87.043002, PhysRevA.107.063301},
while the weak impurity-boson interactions and the large impurity
separations result in the Yukawa potential~\cite{PhysRevX.8.031042,
  doi:10.7566/JPSJ.87.043002, Jager_2022, PhysRevA.107.063301,
  PhysRevA.110.030101}.
Additionally, when the impurity separation exceeds the healing length,
the relativistic van der Waals potential becomes
dominant~\cite{PhysRevLett.129.233401}.

\begin{figure}[b]  
  \centering
  \includegraphics[width=0.47\textwidth]
  {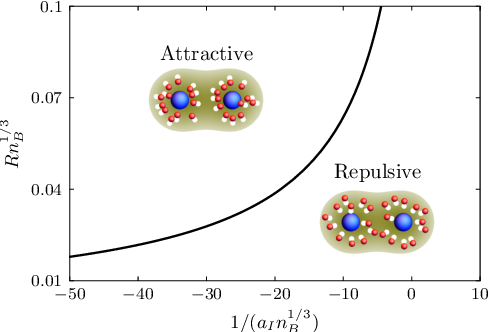}
  \caption{The region of dominant polaron spanned by impurity-boson
    interaction  $1/(a_{I}n_{B}^{1/3})$ and impurities' separation
    $Rn_{B}^{1/3}$.
    The solid line stands for the critical separation distance
    $R_{c}$ as a function of $1/(a_{I}n_{B}^{1/3})$.
    When the impurity-boson interactions are fixed, the excitation is
    more likely an attractive polaron for $R>R_{c}$, while a
    repulsive polaron for $R<R_{c}$.
    In each dominant region, the corresponding polaron carrys more
    spectral weight and damps out more slowly.
    Here we set boson-boson interaction $a_{B}n_{B}^{1/3} = 0.1$.
    The red and blue balls stand for the bosons and impurities,
    respectively.} 
  \label{fig:PhDg_phi_eta_omega}
\end{figure}

In this Letter, the induced interactions between two heavy impurities
are considered from the perspective of Bose polarons.
We show that the ``shifted Newtonian'', Efimov, and Yukawa potentials
can emerge from the attractive polarons under specific conditions.
Intriguingly, when the polarons are repulsive, they are found to be
asymptotically free.
By extending our analysis to both attractive and repulsive polarons
across the Feshbach resonance, we demonstrate that repulsive polarons
can exist even with attractive impurity-boson interactions when the
impurities are sufficiently close, as illustrated in
Fig.~\ref{fig:PhDg_phi_eta_omega}.
These results highlight the interesting double-impurity polarons and
are experimentally testable in cold atomic
systems~\cite{PhysRevA.108.033301}.


\sect{Model}
We consider two stationary impurities immersed in a BEC, which can be
described by the grand canonical Hamiltonian
\begin{eqnarray}
  \label{eq:Hamiltonian}
  H
  &=& \int \ud^{3}\br \left[ \phi^{\dag}
      \left(-\frac{\hbar^{2}\nabla^{2}}{2m_{B}}-\mu\right) 
      \phi + \frac{g_{B}}{2}
      \phi^{\dag}\phi^{\dag}\phi\phi
      \right] \nonumber \\
  & & + \int \ud^{3}\br \phi^{\dag} \left[
      g_{I}\delta(\br-\br_{1}) + g_{I}\delta(\br-\br_{2}) \right]
      \phi,
\end{eqnarray}
where $\phi^{\dag}(\br)$ and $\phi(\br)$ are the creation and
annhilation operators, respectively, for bosonic atoms with mass
$m_{B}$.
Here $\mu$ is the chemical potential,
and $g_{B}=4\pi\hbar^{2}a_{B}/m_{B}$ denotes the boson-boson
interaction with $a_{B}$ being the $s$-wave scattering length.
Two stationary impurities locate at $\br_{1}$ and $\br_{2}$,
and interact with bosonic atom via the contact potential described by
the delta function with the same coupling constant $g_{I}$.
For a strong impurity-boson coupling, $g_{I}$ is related to the
$s$-wave scattering length $a_{I}$ obtained by solving the two-body
Lippmann-Schwinger equation,
$
g_{I}^{-1} = (2\pi\hbar^{2}a_{I}/m_{B})^{-1} -
\sum_{\bq} \epsilon_{\bq}^{-1}/V
$, where
$\epsilon_{\bq} = \hbar^{2}\bq^{2}/2m_{B}$,
and $V$ is the volume of system and
$V^{-1}\sum_{\bq} \rightarrow (2\pi)^{-3}\int \ud^{3}\bq$
at the thermodynamic limit.
For convenience, we shall set $\hbar=1$.

Within the framework of the imaginary-time field integral, we can cast
the partition function of the system as
$
\mathcal{Z} = \int \mathcal{D}[\aphi,\phi]
e^{-S}
$, with the action given by
$
S[\aphi,\phi] = 
\int_{0}^{\beta} \ud\tau \left(
  \int \ud^{3}\br\phi^{\ast}\partial_{\tau}\phi + H[\aphi,\phi]
\right)
$, where $\beta = (k_{B}T)^{-1}$ is the inverse temperature, and
$k_{B}$ represents the Boltzmann constant.
The Fourier transformation of bosonic field is defined by
$\phi(x) = \sum_{p} \phi_{p}e^{ipx} / \sqrt{\beta V}$,
where $x \equiv(\br,\tau)$ and $p \equiv(\bp,\omega_{n})$ satisfying
the inner product $px = \bp\cdot\br - \omega_{n}\tau$ denote the
four-vector and four-momentum, respectively, and
$\omega_{n} = 2n\pi/\beta$ with $n\in\mathbb{Z}$ is the bosonic
Matsubara frequecy.

Applying the Fourier transformation and the Bogoliubov approximation
to the action, we then obtain an effective action
$S \approx
-\frac{1}{2}\beta V g_{B}n_{B}^{2}  + S_{B}
+ \beta\Sigma_{I} + S_{I}$,
where the first two terms is the basic bosonic action in the absence of
impurities, and $n_{B}$ denotes the condensate density.
The second term is the Gaussian action for the bosonic fluctuating
fields and can be compactly written as
$S_{B} =
\frac{1}{2}\sum_{\omega_{n},\bp\neq0}\Phi_{p}^{\dag}[-G_{B}^{-1}(p)]\Phi_{p}$
with the column vector defined as
$\Phi_{p} = (\phi_{p},\aphi_{-p})^{T}$,
and the inverse matrix
$-G_{B}^{-1}(p) = 
-i\omega_{n}\sigma_{z}+ (\epsilon_{\bp}+g_{B}n_{B})I +
g_{B}n_{B}\sigma_{x}$
with the $2 \times 2$ Pauli matrix $\sigma_{x,z}$ and the unit matrix
$I$.
The chemical potential has been taken as $\mu = g_{B}n_{B}$,
which follows from the Hugenholtz-Pines
relation~\cite{PhysRev.116.489} evaluated at the mean-field level.
The last two terms stem from the impurity-boson interactions,
and the mean-field contributions have been grouped into the third term
as 
$
\Sigma_{I} = 2g_{I}\sum_{\omega_{n}} |\phi(0,i\omega_{n})|^{2}/(\beta V)
\approx 2g_{I}n_{B}
$, here we have assumed that the condensate density $n_{B}$ is not
altered by the presence of the impurity \cite{PhysRevA.88.053632}.
The remaining fluctuating components are included in the last term
\begin{subequations}
\begin{align}
  & S_{I} = \frac{g_{I}}{V} \!\!\!
    \sum_{\bp_{1},\bp_{2},\omega_{n}} \!\!\!
    (1-\delta_{\bp_{1}}\delta_{\bp_{2}})    
    U(\bp_{2}-\bp_{1})\aphi_{p_{1}}\phi_{p_{2}}, \\
  & U(\bp_{2}-\bp_{1}) =
    e^{i(\bp_{1}-\bp_{2})\cdot\br_{1}} +
    e^{i(\bp_{1}-\bp_{2})\cdot\br_{2}} ,
\end{align}  
\end{subequations}
where $p_{1,2} = (\bp_{1,2},i\omega_{n})$.

\sect{Wilsonian Renormalization}

At the zero temperature limit, for the weak impurity-boson
interactions, the condensate energy is given by
$E_{G} = \frac{1}{2}Vg_{B}n_{B}^{2} + \Sigma_{I}$.
We note that only $\Sigma_{I}$ is induced by impurities.
For the strong impurity-boson interactions, the $\Sigma_{I}$ can
be obtained by using the Wilsonian renormalization.

\begin{figure}[t]  
  \subfigure{\label{fig:FeynDiag_LI}}
  \subfigure{\label{fig:FeynDiag_Ssq}}
  \subfigure{\label{fig:FeynDiag_Rule}}
  \centering
  \includegraphics[width=0.47\textwidth]{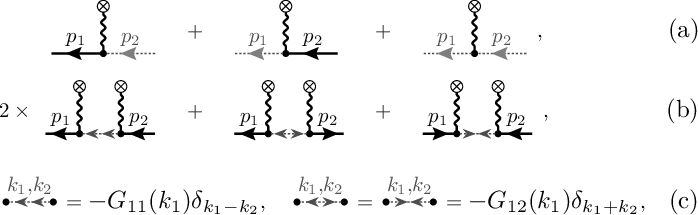}
  \caption{Diagrammatic representation of
    (a) the interaction Lagrangian density $\cL_{I}(p_{1},p_{2})$,
    (b) the slow-mode contributions from
    $\langle \cL_{I}^{2} \rangle_{\f} - \langle \cL_{I}
    \rangle_{\f}^{2}$
    and (c) the fast-mode Feynman rules, with the normal and the
    anomalous propagators given in Eq.~(\ref{eq:GF}).
    The gray dashed (black solid) line denotes the fast (slow) mode.
    While the arrow pointing towards (away from) the vertex stands
    for the annihilation (creation).
    We use the cross dot and a wavy line to represent the
    impurity-boson interaction $g_{I}U(\bp_{1}-\bp_{2})/V$.
  }
  \label{fig:FeynDiag}
\end{figure}

We consider this theory within a momentum cutoff
$\Lambda$, and the effective action can be cast as
\begin{align}
  &S_{\Lambda} = \frac{1}{2}
  \hspace{-0.8em}
  \sum_{\substack{\omega_{n} \\ 0<|\bp|\leqslant\Lambda}}
  \hspace{-0.8em}
  \Phi_{p}^{\dag}[-G_{B}^{-1}(p)]\Phi_{p}
  + \beta\Sigma_{I}(\Lambda) \nonumber \\
  &+ \frac{g_{I}(\Lambda)}{V}
    \hspace{-1.5em}
    \sum_{\substack{\omega_{n} \\ |\bp_{1}|,|\bp_{2}|\leqslant\Lambda}}
    \hspace{-1em}
    (1-\delta_{\bp_{1}}\delta_{\bp_{2}})    
    U(\bp_{2}-\bp_{1})\aphi_{p_{1}}\phi_{p_{2}},
  \label{eq:S_Lambda}
\end{align}
where the irrelevant mean-field contribution is dropped.
Next, we impose a hard cutoff $\Lambda'=\Lambda - \Delta\Lambda$ to
the low-energy theory by decomposing the bosonic fields
$(\aphi_{p},\phi_{p})$ into slow contributions
$(\aphis_{p},\phis_{p})$ and their fast complementary part
$(\aphif_{p},\phif_{p})$, i.e.,
$\phis_{p} = \phi_{p}\Theta(\Lambda'-|\bp|)$ and
$\phif_{p} = \phi_{p}\Theta(|\bp|-\Lambda')$,
where $\Theta(x)$ is the Heaviside step function.
We then divide the action into three parts,
$S_{\Lambda} = S_{\s} + S_{\f} + S_{I}$~\cite{supp},
where the slow part of action $S_{\s}$ is simply given by substituting
all bosonic fields with their slow parts in Eq.~(\ref{eq:S_Lambda}).
And
$ S_{\f} = 
\frac{1}{2}\sum_{\omega_{n},\Lambda'<|\bp|\leqslant\Lambda}
\Phi_{p}^{(\f)\dag}[-G_{B}^{-1}(p)]\Phi^{(\f)}_{p}
$ is the fast part of action, with
$\Phi^{j}_{p} \equiv (\phi^{j}_{p},\phi^{j\ast}_{-p})^{T}$
for $j=(\s)$ or $(\f)$.
And the interaction part of the action is given by
$S_{I} = \sum_{p_{1},p_{2}}\cL_{I}(p_{1},p_{2})$,
where the Lagrangian density $\cL_{I}(p_{1},p_{2})$ is expressed
diagrammatically in Fig.~\ref{fig:FeynDiag_LI}.
To construct an effective theory with the new cutoff
$\Lambda' = \Lambda-\Delta\Lambda$, we need to integrate out all fast
fields, and the effective action is formally given by
$S_{\Lambda'} \approx S_{\s} + \langle S_{I} \rangle_{\f}
- \frac{1}{2} \left(
  \langle S_{I}^{2} \rangle_{\f} - \langle S_{I} \rangle_{\f}^{2}
\right)$, where the fast-mode average of $X$ is defined by
$ \langle X \rangle_{\f} = \int\cD[\aphif,\phif]Xe^{-S_{\f}} /
\int\cD[\aphif,\phif]e^{-S_{\f}}$.
The slow-mode contribution from
$\langle S_{I} \rangle_{\f} = \sum_{p_{1},p_{2}} \langle \cL_{I}
\rangle_{\f}$ is found to be zero.
While the slow-mode contributions from
$\langle S_{I}^{2} \rangle_{\f} - \langle S_{I} \rangle_{\f}^{2}$
are shown in Fig.~\ref{fig:FeynDiag_Ssq}.

The Feynman rules are given in Fig.~\ref{fig:FeynDiag_Rule}, where
the normal and the anomalous propagators are respectively given by
\begin{subequations}
  \label{eq:GF}
\begin{align}
  & G_{11}(k) = G_{11}(\bk,i\omega_{n})
    = \frac{i\omega_{n} + \epsilon_{\bk} + g_{B}n_{B}}
    {(i\omega_{n})^{2}-\omega_{B}^{2}(\bk)}, \\
  & G_{12}(k) = G_{12}(\bk,i\omega_{n})
    = -\frac{g_{B}n_{B}}
    {(i\omega_{n})^{2}-\omega_{B}^{2}(\bk)},    
\end{align}  
\end{subequations}
with
$\omega_{B}(\bk) = \sqrt{\bk^{2}
  (\bk^{2}+4m_{B}g_{B}n_{B})}/{(2m_{B})}$
being the Bogoliubov excitation.
Collecting and rearranging all slow-mode contributions yields the
effective action $S_{\Lambda'}$, which has the same form as
Eq.~(\ref{eq:S_Lambda}) but with the new lower cutoff
$\Lambda' = \Lambda-\Delta\Lambda$.

Under this coarse-graining transformation,
$S_{\Lambda} \rightarrow S_{\Lambda'}$,
the impurity-boson coupling and the mean-field energy contribution are
renormalized as $g_{I}(\Lambda) \rightarrow g_{I}(\Lambda')$ and
$\Sigma_{I}(\Lambda) \rightarrow \Sigma_{I}(\Lambda')$, respectively.
The mean-field energy contributions to $\Sigma_{I}(\Lambda')$ are
obtained by setting the external momentum-energy of
Fig.~\ref{fig:FeynDiag_Ssq} to condensate mode, i.e.,
$(\bp,i\omega) \rightarrow (0,\omega_{I})$,
where $\omega_{I}$ is the induced energy for $\Lambda \rightarrow 0$,
that is
$\omega_{I} = \Sigma_{I}(\Lambda\rightarrow0)$.
As the $\Delta\Lambda$ becoming infinitesimal, the renormalization
transformation yields the flow equation~\cite{supp}
\begin{align}
  \hspace{-0.5em}
  \frac{\ud\Sigma_{I}}{\ud\Lambda} 
  \!\approx\!
  -\frac{\Sigma_{I}^{2}\Lambda^{2}}{2n_{B}(2\pi)^{3}}
  \!\int\!
  \ud\Omega \left( 1+e^{i\Lambda R \cos\theta} \right)
  \! G(\Lambda,\omega_{I}),
  \hspace{-0.5em}
\end{align}
where $\ud\Omega = \sin\theta\ud\theta\ud\varphi$ is the differential
solid angle, $R=|\bR|=|\br_{1}-\br_{2}|$ is the distance between two
impurities, and
$G(\bk,\omega) = G_{11}(\bk,\omega) +
G_{12}(\bk,\omega)$.
Here we have taken the advantage of
$\Sigma_{I}(\Lambda) \approx 2g_{I}(\Lambda)n_{B}$.
With the renormalization flow equation in hand, we deduce that the
condensate energy induced by impurity-boson interactions is given by
\begin{align}
  \hspace{-0.5em}
  \frac{1}{\Sigma_{I}(\omega)}
  \!=\!
  \frac{1}{\Sigma_{I}(\Lambda)} \!-\! \frac{1}{2n_{B}V}\!
  \sum_{|\bq|\leqslant\Lambda}\! \left( 1+e^{i\bq\cdot\bR}\right)\!
  G(\bq,\omega),
  \hspace{-0.5em}
  \label{eq:Sigma_Origin}
\end{align}
where
$\Sigma_{I}^{-1}(\Lambda) = [ (2\pi a_{I}/m_{B})^{-1} -
\sum_{\Lambda} \epsilon_{\bq}^{-1}/V ]/(2n_{B})$.
For convenience, we have denoted
$\Sigma_{I}(\omega) = \Sigma_{I}(\Lambda\rightarrow0)$,
which is actually a function of energy $\omega$,
the distance $R$ as well as the scattering length $a_{I}$ and $a_{B}$.

Notice that, by taking transformations $\bR\rightarrow0$,
$2g_{I}\rightarrow g_{I}$ and excluding the contributions of
$G_{12}(q)$, Eq.~(\ref{eq:Sigma_Origin}) will eventually accord with
the results obtained from Ref.~\cite{PhysRevA.88.053632} by neglecting
the depletion-dressing term within the non-self-consistent $T$-matrix
and the Born-Oppenheimer approximation. That is, the self-energy of
double impurities could be reduced to the single one~\cite{supp}.
And fortunately, we find that 
$\Sigma_{I}(\omega+i0^{+}) = \rRe\Sigma_{I}(\omega) +
i\rIm\Sigma_{I}(\omega)$ can be determined analytically~\cite{supp}.


\sect{Effective Interactions}

The ground-state energy of condensate is given by
$E_{G} = \frac{1}{2}Vg_{B}n_{B}^{2} + \omega_{I}$, where
the impurity-induced energy $\omega_{I}$
can be determined by seeking the solutions of equation
\begin{equation}
  \omega_{I} = \rRe\Sigma_{I}(\omega_{I}).
  \label{eq:Pol_Energy}
\end{equation}
The effective interaction between two impurities $V(R)$ is then given
by $V(R) = \omega_{I}(R)-\omega_{I}(R\rightarrow\infty)$, provided
that $\omega_{I}<0$.
We will now proceed to examine the effective interactions under
various conditions.

Firstly, we consider the case of an ideal BEC, where the boson-boson
interaction vanishes as $g_{B}n_{B} \rightarrow 0$. 
To prevent the collapse of BEC, the ground-state energy should be
finite and small, so that Eq.~(\ref{eq:Pol_Energy}) reduces to
$\omega_{I} = \rRe\Sigma_{I}(0)$, which leads to the known
solution~\cite{Panochko_2021, supp},
$\omega_{BEC} = (4\pi n_{B}/m_{B})/ (a_{I}^{-1} + R^{-1})$
with $a_{I}<0$ and $R>|a_{I}|$,
and hence yields a ``shifted Newtonian'' attractive
potential~\cite{PhysRevA.107.063301},
$V_{SN}(R) = -(4\pi n_{B}a_{I}^{2}/m_{B})/ (a_{I} + R)$.

Next, we consider the collapse of BEC, which
can be regarded as all bosons entering the bound
state~\cite{PhysRevA.107.063301}.
In this case, either ideal or weakly interacting BEC should exhibit
$|\omega_{I}|\gg g_{B}n_{B}$ and $\omega_{I}<0$.
Hence, in the unit of $g_{B}n_{B}$, Eq.~(\ref{eq:Pol_Energy})
approaches $[\rRe\Sigma_{I}(\omega_{I})]^{-1}=0$, which leads to a
three-body binding energy for $a_{I}^{-1}>-R^{-1}$, 
\begin{equation}
  \omega_{b} = -\frac{1}{2m_{B}} \left[
    \frac{1}{a_{I}} + \frac{1}{R}
    W\left(e^{-R/a_{I}}\right)
  \right]^{2},
  \label{eq:Three-Body_Binding_Energy}
\end{equation}
where $W(x)$ is the Lambert $W$ function~\cite{PhysRevA.79.013629,
  PhysRevA.102.063321, PhysRevA.107.063301, supp}.
Hence in the large scattering length limit
$1/(a_{I}n_{B}^{1/3}) \rightarrow 0$ and at short distance
$Rn_{B}^{1/3} \ll 1$,
Eq.~(\ref{eq:Three-Body_Binding_Energy}) yields the Efimov
attraction~\cite{doi:10.7566/JPSJ.87.043002, supp},
$V_{E}(R) = -W^{2}(1)/(2m_{B}R^{2})$.

On the other hand, let us consider the case of weakly interacting
impurity, $|\omega_{I}|\ll g_{B}n_{B}$.
In this case, the impurity-induced energy can be approximated
as~\cite{atoms10010019, supp}
\begin{equation}
  \omega_{\mathrm{Weak}} \!=\!
  \frac{4\pi n_{B}^{2/3}/m_{B}}{
    1/(a_{I}n_{B}^{1/3}) -
    \lambda\left[\sqrt{2}+V_{Y}(R)\right]
  },
  \label{eq:Weak_Int_Energy}
\end{equation}
where $\lambda = \sqrt{8\pi a_{B}n_{B}^{1/3}}$,
$V_{Y}(R)=-e^{-\sqrt{2}R/\xi}/(R/\xi)$ 
with $\xi=1/(\lambda n_{B}^{1/3})$ being the healing length.
And if $R/\xi \gg 1$ and $|a_{I}n_{B}^{1/3}|\ll1$,
then Eq.~(\ref{eq:Weak_Int_Energy}) yields an effective Yukawa
potential $\propto V_{Y}(R)$ between two impurities, which is a
well-known result shown in Refs.~\cite{PhysRevX.8.031042,
  doi:10.7566/JPSJ.87.043002, Jager_2022, PhysRevA.107.063301, supp}.

\sect{Polaron Spectrum across Feshbach Resonance}

\begin{figure}[t]  
  \centering
  \includegraphics[width=0.47\textwidth]
  {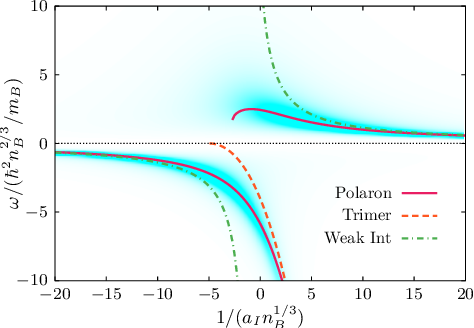}
  \caption{The double-impurity polaron spectrum throughout the
    attractive-repulsive crossover for separation distance
    $Rn_{B}^{1/3} = 0.2$.
    The more deeply shaded region indicates the higher spectral
    weight.
    Solid line: excitation energy obtained by solving
    Eq.~(\ref{eq:Pol_Energy}).
    Dashed line: binding energy of impurity-boson-impurity trimer
    given by Eq.~(\ref{eq:Three-Body_Binding_Energy}).
    Dash-dotted line: energy in the weak-coupling approximation
    obtained by Eq.~(\ref{eq:Weak_Int_Energy}).
    A negative (positive) excitation corresponds to an attractive
    (repulsive) polaron.
    Here $a_{B}n_{B}^{1/3}=0.1$.}
  \label{fig:Spec_Fun_A_Two_Imp}
\end{figure}

For a fixed separation distance $R$, we show the spectrum of the
double-impurity polaron in Fig.~\ref{fig:Spec_Fun_A_Two_Imp},
where the spectral function is given by
\begin{align}
  A(\omega,R) = -2\rIm \left[
  \frac{1}{\omega-\Sigma_{I}(\omega+i0^{+})+i0^{+}}
  \right].
\end{align}
And to respect the condition of dilute gases, without loss of
generality, we set $a_{B}n_{B}^{1/3}=0.1$ hereinafter.

Compared with the polaron spectrum in
Ref.~\cite{PhysRevA.88.053632}, this double-impurity spectrum has the
same overall structure but is slightly shifted to the left.
When crossing the Feshbach resonance, there still exists a crossover
of polaron from attractive to repulsive.
In the weak-coupling regime, both attractive and repulsive
polarons obey Eq.~(\ref{eq:Weak_Int_Energy}), which still approaches
$4\pi a_{I}n_{B}/m_{B}$ at the weak-coupling limit as expected.
However, when the negative excitation becomes large, the attractive
branch of polaron approaches the energy of trimer
Eq.~(\ref{eq:Three-Body_Binding_Energy}) instead of dimer.

\begin{figure}[t]  
  \centering
  \includegraphics[width=0.47\textwidth]
  {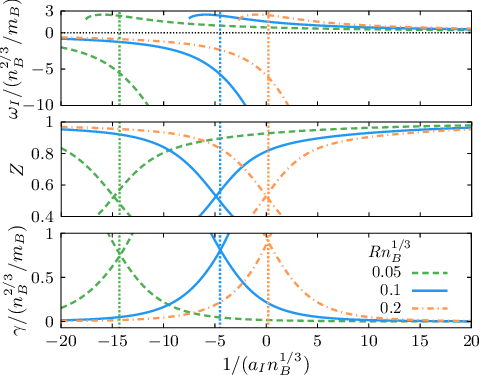}
  \caption{Polaron excitation $\omega_{I}$ (in unit of
    $n_{B}^{2/3}/m_{B}$), spectral weight $Z(\omega_{I})$
    (dimensionless) and damping rate $\gamma(\omega_{I})$ (in unit of
    $n_{B}^{2/3}/m_{B}$) for $Rn_{B}^{1/3} = 0.05$ (dashed line),
    $0.1$ (solid line) and $0.2$ (dash-dotted line).
    The vertical dotted lines mark the points where the damping rates
    of these two polaron branches (attractive and repulsive) converge.
    Here $a_{B}n_{B}^{1/3} = 0.1$.}
  \label{fig:Pol_ExcSpec}
\end{figure}

The key difference with the double-impurity polaron is that the closer
two impurities are, the more spectrum shifts to the left, as shown
in Fig.~\ref{fig:Pol_ExcSpec}.
To quantify the shift, here we also extract the spectral
weight $Z(\omega_{I})$ and damping rate $\gamma(\omega_{I})$ from the
self-energy $\Sigma_{I}(\omega_{I}+i0^{+})$ by
\begin{align}
  & Z(\omega_{I}) =
    \left. \frac{1}{1-\partial_{\omega}\rRe\Sigma_{I}(\omega)}
    \right|_{\omega=\omega_{I}}, \\ 
  & \gamma(\omega_{I}) = -Z(\omega_{I})\rIm\Sigma_{I}(\omega_{I}).
\end{align}
Note that, for a large separation distance $Rn_{B}^{1/3}$,
our results of the polaron excitation $\omega_{I}$ and damping rate
$\gamma(\omega_{I})$ qualitatively coincide with the experimental
results in Ref.~\cite{PhysRevLett.117.055302},
since the double-impurity polaron in this case can be regarded as the
simple superposition of two single-impurity polarons.

As $1/(a_{I}n_{B}^{1/3})$ crosses the Feshbach resonance from negative
side, the attractive polaron loses spectral weight and damps out more
quickly. 
Meanwhile, the repulsive polaron gains spectral weight and damps out
more slowly.
The spectral weights or damping rates of these two polaron branches
eventually converge, marking the crossover from attractive- to
repulsive-polaron domination, as the vertical dotted lines imply in
Fig.~\ref{fig:Pol_ExcSpec}.
Although the crossover points of the spectral weight and damping rate
for each $Rn_{B}^{1/3}$ are slightly different, 
we choose the intersection of damping rates to specify the
attractive-repulsive crossover, which is based on the fact that
damping rate can not only characterize the excitation lifetime but
also be measured experimentally~\cite{PhysRevLett.117.055301,
  PhysRevLett.117.055302}.

Since the crossover point shifts towards the negative direction of
$1/(a_{I}n_{B}^{1/3})$ as impurities getting closer, it is surprising
that the repulsive polaron can stably exist over attractive polaron
even when the impurity-boson interaction is attractive.
This is quite different from the single-impurity case, where the
repulsive polaron is dominant only for repulsive impurity-boson
interaction~\cite{PhysRevLett.117.055301, PhysRevLett.117.055302,
  PhysRevA.88.053632}.

\sect{Asymptotic Freedom}

\begin{figure}[t]  
  \centering
  \includegraphics[width=0.47\textwidth]
  {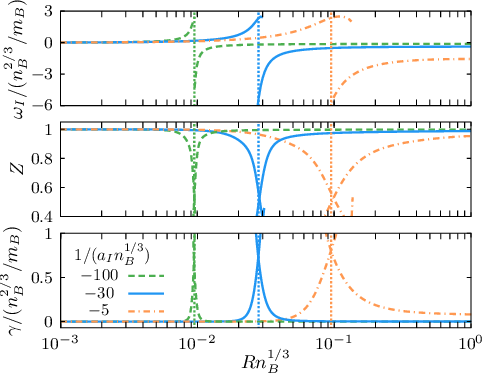}
  \caption{Polaron excitation $\omega_{I}$ (in unit of
    $n_{B}^{2/3}/m_{B}$), spectral weight $Z(\omega_{I})$
    (dimensionless) and damping rate $\gamma(\omega_{I})$ (in unit of
    $n_{B}^{2/3}/m_{B}$) for $1/(a_{I}n_{B}^{1/3}) = -100$ (dashed line),
    $-30$ (solid line) and $-5$ (dash-dotted line).
    The vertical dotted lines mark the critical separation distance
    $R_{c}$ where the damping rates of two polaron are equal.
    Here $a_{B}n_{B}^{1/3} = 0.1$.}
  \label{fig:Pol_ExcSpec_R}
\end{figure}

This double-impurity polaron can be viewed from another perspective.
We now fix the impurity-boson interaction $a_{I}$,
and consider the excitation energy $\omega_{I}$ varying with the
separation distance $R$. 
If the above arguments are valid, one will find that there is a
critical separation distance $R_{c}$ which delimits the dominant
boundary of attractive and repulsive polaron for $a_{I}<0$.
In Fig.~\ref{fig:Pol_ExcSpec_R}, we use the vertical dotted lines to
mark out this critical distance. 
When the boson-boson interaction ($a_{B}n_{B}^{1/3} = 0.1$) is fixed,
$R_{c}n_{B}^{1/3}$ is found to be monotonically increasing as
$1/(a_{I}n_{B}^{1/3})$ approaches the resonance for $a_{I}<0$,
as shown in Fig.~\ref{fig:PhDg_phi_eta_omega}.

When $R>R_{c}$, the attractive polaron carrys more spectral weight and
damps out more slowly, so that the attractive polaron becomes dominant
in this region.
As the separation distance $R$ increases, the excitation energy
$\omega_{I}$ approaches twice that of the single-impurity polaron,
which is expected to be $4\pi a_{I}n_{B}/m_{B}$ when
$|a_{I}n_{B}^{1/3}|\ll1$.
In contrast, for $R<R_{c}$, spectral weight $Z$ approaching $1$ and
damping rate $\gamma$ vanishing rapidly in the limit of short
separation distance, indicates that the excitation is more likely a
repulsive polaron.
The existence of repulsive dominant polaron for $a_{I}<0$,
which is distinct from the case of single-impurity polaron,
can be understood from the quantum blocking
effect~\cite{PhysRevLett.127.033401, PhysRevLett.121.243401,
  PhysRevA.98.062705}:
when two impurities become close enough, an impurity-boson-impurity
trimer will form and block the bosons nearby, so that the whole trimer
can be regarded as an impurity repelling other bosons.
Since the trimer energy can be characterized by
Eq.~(\ref{eq:Three-Body_Binding_Energy}), one can deduce that
impurities with weaker attractive interactions need a smaller distance
$R$ to reach the same binding energy.
This implies that the trimer size, hence the repulsive region,
decreases as the attractive impurity-boson interactions are tuned
down, which qualitatively explains the
Fig.~\ref{fig:PhDg_phi_eta_omega}.

In the repulsive-polaron regime, as two impurities get closer,
the excitation energy tends to vanish,
and these two impurities are asymptotically free.
In fact, the feature of asymptotic freedom exists regardless of
$a_{I}$ being.
The reason is that: if two impurities are close enough,
the weak energy $\omega_{\mathrm{Weak}}$ in
Eq.~(\ref{eq:Weak_Int_Energy}) will become positive, and it is so
small that can be regarded as the actual polaron excitation.
And the asymptotic freedom is guaranteed by
$\omega_{I} \approx \omega_{\mathrm{Weak}} \rightarrow 0$ as
$R \rightarrow 0$.
In fact, for a weak repulsive impurity-boson interaction that
satisfies $1/(a_{I}n_{B}^{1/3}) \gg \sqrt{16\pi a_{B}n_{B}^{1/3}}$,
$\omega_{I} \approx \omega_{\mathrm{Weak}}$ holds for all separation
distance $R$.
And $\omega_{I}$ linearly depends on $R$ in the short-distance limit
as $\omega_{\mathrm{Weak}} \approx 4\pi n_{B}R/m_{B}$~\cite{supp}.


Based on the above results, we argue that the effective interaction
between impurities depends on the dominant polaron.
To clarify, we consider the impurity-boson interaction to be weak.
In the attractive-polaron regime, the effective interaction did change
from Yukawa- to Efimov-type as impurities approach each other from a
large separation~\cite{PhysRevA.107.063301}.
In the repulsive-polaron regime, the effective interaction, given by
Eq.~(\ref{eq:Weak_Int_Energy}), changes from Yukawa-type to linear as
impurities approach, and vanishes in the asymptotic limit.

In principle, the heavy impurities immersed in a BEC can be realized
by using $^{7}$Li and $^{133}$Cr~\cite{PhysRevA.108.033301}. 
The separation distance of impurities can be manipulated by optical
tweezer~\cite{Kaufman2021} directly, or by varying the condensate
density while fixing impurities.
The energy and damping rate can be characterized by using the
radio-frequency
spectroscopy~\cite{PhysRevLett.117.055301, PhysRevLett.117.055302}.

\sect{Conclusion}

In conclusion, we consider two static heavy impurities immersed in a
BEC, and calculate the self-energy using the Wilsonian
renormalization.
From the self-energy, we extract the polaron energy, quasiparticle
residue and damping rate, which depend not only on the impurity-boson
interactions but also on the separation distance between the
impurities.
The outcomes are solid and able to reproduce many
established results.
We then determine the attractive-repulsive crossover boundary of 
polaron based on the damping rate. 
Our analysis reveals that repulsive-dominant polarons exist as long
as the impurities are sufficiently close, even if the impurity-boson
interactions are attractive.
Furthermore, we observe that the two impurities become asymptotically
free in the repulsive polaron regime.
Our findings provide a new perspective on the interaction between two
polarons, which can be directly tested in cold-atom experiments.

\sect{Acknowledgments}
\begin{acknowledgments}
  R.L. acknowledges the funding from the NSFC under Grants No.12174055
  and No.11674058.
  L.W. acknowledges the funding from the NSFC under Grants No.12175027
  and No.11875010.
\end{acknowledgments}


\begin{thebibliography}{55}%
\makeatletter
\providecommand \@ifxundefined [1]{%
 \@ifx{#1\undefined}
}%
\providecommand \@ifnum [1]{%
 \ifnum #1\expandafter \@firstoftwo
 \else \expandafter \@secondoftwo
 \fi
}%
\providecommand \@ifx [1]{%
 \ifx #1\expandafter \@firstoftwo
 \else \expandafter \@secondoftwo
 \fi
}%
\providecommand \natexlab [1]{#1}%
\providecommand \enquote  [1]{``#1''}%
\providecommand \bibnamefont  [1]{#1}%
\providecommand \bibfnamefont [1]{#1}%
\providecommand \citenamefont [1]{#1}%
\providecommand \href@noop [0]{\@secondoftwo}%
\providecommand \href [0]{\begingroup \@sanitize@url \@href}%
\providecommand \@href[1]{\@@startlink{#1}\@@href}%
\providecommand \@@href[1]{\endgroup#1\@@endlink}%
\providecommand \@sanitize@url [0]{\catcode `\\12\catcode `\$12\catcode
  `\&12\catcode `\#12\catcode `\^12\catcode `\_12\catcode `\%12\relax}%
\providecommand \@@startlink[1]{}%
\providecommand \@@endlink[0]{}%
\providecommand \url  [0]{\begingroup\@sanitize@url \@url }%
\providecommand \@url [1]{\endgroup\@href {#1}{\urlprefix }}%
\providecommand \urlprefix  [0]{URL }%
\providecommand \Eprint [0]{\href }%
\providecommand \doibase [0]{https://doi.org/}%
\providecommand \selectlanguage [0]{\@gobble}%
\providecommand \bibinfo  [0]{\@secondoftwo}%
\providecommand \bibfield  [0]{\@secondoftwo}%
\providecommand \translation [1]{[#1]}%
\providecommand \BibitemOpen [0]{}%
\providecommand \bibitemStop [0]{}%
\providecommand \bibitemNoStop [0]{.\EOS\space}%
\providecommand \EOS [0]{\spacefactor3000\relax}%
\providecommand \BibitemShut  [1]{\csname bibitem#1\endcsname}%
\let\auto@bib@innerbib\@empty
\bibitem [{\citenamefont {Yukawa}(1935)}]{10.1143/PTPS.1.1}%
  \BibitemOpen
  \bibfield  {author} {\bibinfo {author} {\bibfnamefont {H.}~\bibnamefont
  {Yukawa}},\ }\bibfield  {title} {\bibinfo {title} {{On the Interaction of
  Elementary Particles. I}},\ }\href {https://doi.org/10.1143/PTPS.1.1}
  {\bibfield  {journal} {\bibinfo  {journal} {Proc. Phys. Math. Soc. Jpn.}\
  }\textbf {\bibinfo {volume} {17}},\ \bibinfo {pages} {48} (\bibinfo {year}
  {1935})}\BibitemShut {NoStop}%
\bibitem [{\citenamefont {Dirac}\ and\ \citenamefont
  {Bohr}(1927)}]{doi:10.1098/rspa.1927.0039}%
  \BibitemOpen
  \bibfield  {author} {\bibinfo {author} {\bibfnamefont {P.~A.~M.}\
  \bibnamefont {Dirac}}\ and\ \bibinfo {author} {\bibfnamefont {N.~H.~D.}\
  \bibnamefont {Bohr}},\ }\bibfield  {title} {\bibinfo {title} {{The quantum
  theory of the emission and absorption of radiation}},\ }\href
  {https://doi.org/10.1098/rspa.1927.0039} {\bibfield  {journal} {\bibinfo
  {journal} {Proc. R. Soc. A}\ }\textbf {\bibinfo {volume} {114}},\ \bibinfo
  {pages} {243} (\bibinfo {year} {1927})}\BibitemShut {NoStop}%
\bibitem [{\citenamefont {Yang}\ and\ \citenamefont
  {Mills}(1954)}]{PhysRev.96.191}%
  \BibitemOpen
  \bibfield  {author} {\bibinfo {author} {\bibfnamefont {C.~N.}\ \bibnamefont
  {Yang}}\ and\ \bibinfo {author} {\bibfnamefont {R.~L.}\ \bibnamefont
  {Mills}},\ }\bibfield  {title} {\bibinfo {title} {{Conservation of Isotopic
  Spin and Isotopic Gauge Invariance}},\ }\href
  {https://doi.org/10.1103/PhysRev.96.191} {\bibfield  {journal} {\bibinfo
  {journal} {Phys. Rev.}\ }\textbf {\bibinfo {volume} {96}},\ \bibinfo {pages}
  {191} (\bibinfo {year} {1954})}\BibitemShut {NoStop}%
\bibitem [{\citenamefont {Gross}\ and\ \citenamefont
  {Wilczek}(1973)}]{PhysRevLett.30.1343}%
  \BibitemOpen
  \bibfield  {author} {\bibinfo {author} {\bibfnamefont {D.~J.}\ \bibnamefont
  {Gross}}\ and\ \bibinfo {author} {\bibfnamefont {F.}~\bibnamefont
  {Wilczek}},\ }\bibfield  {title} {\bibinfo {title} {{Ultraviolet Behavior of
  Non-Abelian Gauge Theories}},\ }\href
  {https://doi.org/10.1103/PhysRevLett.30.1343} {\bibfield  {journal} {\bibinfo
   {journal} {Phys. Rev. Lett.}\ }\textbf {\bibinfo {volume} {30}},\ \bibinfo
  {pages} {1343} (\bibinfo {year} {1973})}\BibitemShut {NoStop}%
\bibitem [{\citenamefont {Politzer}(1973)}]{PhysRevLett.30.1346}%
  \BibitemOpen
  \bibfield  {author} {\bibinfo {author} {\bibfnamefont {H.~D.}\ \bibnamefont
  {Politzer}},\ }\bibfield  {title} {\bibinfo {title} {{Reliable Perturbative
  Results for Strong Interactions?}},\ }\href
  {https://doi.org/10.1103/PhysRevLett.30.1346} {\bibfield  {journal} {\bibinfo
   {journal} {Phys. Rev. Lett.}\ }\textbf {\bibinfo {volume} {30}},\ \bibinfo
  {pages} {1346} (\bibinfo {year} {1973})}\BibitemShut {NoStop}%
\bibitem [{\citenamefont {Bardeen}\ \emph {et~al.}(1957)\citenamefont
  {Bardeen}, \citenamefont {Cooper},\ and\ \citenamefont
  {Schrieffer}}]{PhysRev.108.1175}%
  \BibitemOpen
  \bibfield  {author} {\bibinfo {author} {\bibfnamefont {J.}~\bibnamefont
  {Bardeen}}, \bibinfo {author} {\bibfnamefont {L.~N.}\ \bibnamefont
  {Cooper}},\ and\ \bibinfo {author} {\bibfnamefont {J.~R.}\ \bibnamefont
  {Schrieffer}},\ }\bibfield  {title} {\bibinfo {title} {{Theory of
  Superconductivity}},\ }\href {https://doi.org/10.1103/PhysRev.108.1175}
  {\bibfield  {journal} {\bibinfo  {journal} {Phys. Rev.}\ }\textbf {\bibinfo
  {volume} {108}},\ \bibinfo {pages} {1175} (\bibinfo {year}
  {1957})}\BibitemShut {NoStop}%
\bibitem [{\citenamefont {Landau}\ and\ \citenamefont
  {Pekar}(1948)}]{landau1948effective}%
  \BibitemOpen
  \bibfield  {author} {\bibinfo {author} {\bibfnamefont {L.}~\bibnamefont
  {Landau}}\ and\ \bibinfo {author} {\bibfnamefont {S.}~\bibnamefont {Pekar}},\
  }\bibfield  {title} {\bibinfo {title} {{Effective mass of a polaron}},\
  }\href {http://archive.ujp.bitp.kiev.ua/files/journals/53/si/53SI15p.pdf}
  {\bibfield  {journal} {\bibinfo  {journal} {Zh. Eksp. Teor. Fiz}\ }\textbf
  {\bibinfo {volume} {18}},\ \bibinfo {pages} {419} (\bibinfo {year}
  {1948})}\BibitemShut {NoStop}%
\bibitem [{\citenamefont {Bloch}\ \emph {et~al.}(2008)\citenamefont {Bloch},
  \citenamefont {Dalibard},\ and\ \citenamefont {Zwerger}}]{RevModPhys.80.885}%
  \BibitemOpen
  \bibfield  {author} {\bibinfo {author} {\bibfnamefont {I.}~\bibnamefont
  {Bloch}}, \bibinfo {author} {\bibfnamefont {J.}~\bibnamefont {Dalibard}},\
  and\ \bibinfo {author} {\bibfnamefont {W.}~\bibnamefont {Zwerger}},\
  }\bibfield  {title} {\bibinfo {title} {{Many-body physics with ultracold
  gases}},\ }\href {https://doi.org/10.1103/RevModPhys.80.885} {\bibfield
  {journal} {\bibinfo  {journal} {Rev. Mod. Phys.}\ }\textbf {\bibinfo {volume}
  {80}},\ \bibinfo {pages} {885} (\bibinfo {year} {2008})}\BibitemShut
  {NoStop}%
\bibitem [{\citenamefont {Grusdt}\ \emph {et~al.}(2024)\citenamefont {Grusdt},
  \citenamefont {Mostaan}, \citenamefont {Demler},\ and\ \citenamefont
  {Ardila}}]{grusdt2024impuritiespolaronsbosonicquantum}%
  \BibitemOpen
  \bibfield  {author} {\bibinfo {author} {\bibfnamefont {F.}~\bibnamefont
  {Grusdt}}, \bibinfo {author} {\bibfnamefont {N.}~\bibnamefont {Mostaan}},
  \bibinfo {author} {\bibfnamefont {E.}~\bibnamefont {Demler}},\ and\ \bibinfo
  {author} {\bibfnamefont {L.~A.~P.}\ \bibnamefont {Ardila}},\ }\href
  {https://arxiv.org/abs/2410.09413} {\bibinfo {title} {{Impurities and
  polarons in bosonic quantum gases: a review on recent progress}}} (\bibinfo
  {year} {2024}),\ \Eprint {https://arxiv.org/abs/2410.09413} {arXiv:2410.09413
  [cond-mat.quant-gas]} \BibitemShut {NoStop}%
\bibitem [{\citenamefont {Baroni}\ \emph
  {et~al.}(2024{\natexlab{a}})\citenamefont {Baroni}, \citenamefont
  {Lamporesi},\ and\ \citenamefont {Zaccanti}}]{Baroni2024NatRevPhys}%
  \BibitemOpen
  \bibfield  {author} {\bibinfo {author} {\bibfnamefont {C.}~\bibnamefont
  {Baroni}}, \bibinfo {author} {\bibfnamefont {G.}~\bibnamefont {Lamporesi}},\
  and\ \bibinfo {author} {\bibfnamefont {M.}~\bibnamefont {Zaccanti}},\
  }\bibfield  {title} {\bibinfo {title} {{Quantum mixtures of ultracold gases
  of neutral atoms}},\ }\href {https://doi.org/10.1038/s42254-024-00773-6}
  {\bibfield  {journal} {\bibinfo  {journal} {Nat Rev Phys}\ }\textbf {\bibinfo
  {volume} {6}},\ \bibinfo {pages} {736} (\bibinfo {year}
  {2024}{\natexlab{a}})}\BibitemShut {NoStop}%
\bibitem [{\citenamefont {Baroni}\ \emph
  {et~al.}(2024{\natexlab{b}})\citenamefont {Baroni}, \citenamefont {Huang},
  \citenamefont {Fritsche}, \citenamefont {Dobler}, \citenamefont {Anich},
  \citenamefont {Kirilov}, \citenamefont {Grimm}, \citenamefont
  {Bastarrachea-Magnani}, \citenamefont {Massignan},\ and\ \citenamefont
  {Bruun}}]{Baroni2024NatPhys}%
  \BibitemOpen
  \bibfield  {author} {\bibinfo {author} {\bibfnamefont {C.}~\bibnamefont
  {Baroni}}, \bibinfo {author} {\bibfnamefont {B.}~\bibnamefont {Huang}},
  \bibinfo {author} {\bibfnamefont {I.}~\bibnamefont {Fritsche}}, \bibinfo
  {author} {\bibfnamefont {E.}~\bibnamefont {Dobler}}, \bibinfo {author}
  {\bibfnamefont {G.}~\bibnamefont {Anich}}, \bibinfo {author} {\bibfnamefont
  {E.}~\bibnamefont {Kirilov}}, \bibinfo {author} {\bibfnamefont
  {R.}~\bibnamefont {Grimm}}, \bibinfo {author} {\bibfnamefont {M.~A.}\
  \bibnamefont {Bastarrachea-Magnani}}, \bibinfo {author} {\bibfnamefont
  {P.}~\bibnamefont {Massignan}},\ and\ \bibinfo {author} {\bibfnamefont
  {G.~M.}\ \bibnamefont {Bruun}},\ }\bibfield  {title} {\bibinfo {title}
  {{Mediated interactions between Fermi polarons and the role of impurity
  quantum statistics}},\ }\href {https://doi.org/10.1038/s41567-023-02248-4}
  {\bibfield  {journal} {\bibinfo  {journal} {Nat. Phys.}\ }\textbf {\bibinfo
  {volume} {20}},\ \bibinfo {pages} {68} (\bibinfo {year}
  {2024}{\natexlab{b}})}\BibitemShut {NoStop}%
\bibitem [{\citenamefont {Schirotzek}\ \emph {et~al.}(2009)\citenamefont
  {Schirotzek}, \citenamefont {Wu}, \citenamefont {Sommer},\ and\ \citenamefont
  {Zwierlein}}]{PhysRevLett.102.230402}%
  \BibitemOpen
  \bibfield  {author} {\bibinfo {author} {\bibfnamefont {A.}~\bibnamefont
  {Schirotzek}}, \bibinfo {author} {\bibfnamefont {C.-H.}\ \bibnamefont {Wu}},
  \bibinfo {author} {\bibfnamefont {A.}~\bibnamefont {Sommer}},\ and\ \bibinfo
  {author} {\bibfnamefont {M.~W.}\ \bibnamefont {Zwierlein}},\ }\bibfield
  {title} {\bibinfo {title} {{Observation of Fermi Polarons in a Tunable Fermi
  Liquid of Ultracold Atoms}},\ }\href
  {https://doi.org/10.1103/PhysRevLett.102.230402} {\bibfield  {journal}
  {\bibinfo  {journal} {Phys. Rev. Lett.}\ }\textbf {\bibinfo {volume} {102}},\
  \bibinfo {pages} {230402} (\bibinfo {year} {2009})}\BibitemShut {NoStop}%
\bibitem [{\citenamefont {Nascimb\`ene}\ \emph {et~al.}(2009)\citenamefont
  {Nascimb\`ene}, \citenamefont {Navon}, \citenamefont {Jiang}, \citenamefont
  {Tarruell}, \citenamefont {Teichmann}, \citenamefont {McKeever},
  \citenamefont {Chevy},\ and\ \citenamefont
  {Salomon}}]{PhysRevLett.103.170402}%
  \BibitemOpen
  \bibfield  {author} {\bibinfo {author} {\bibfnamefont {S.}~\bibnamefont
  {Nascimb\`ene}}, \bibinfo {author} {\bibfnamefont {N.}~\bibnamefont {Navon}},
  \bibinfo {author} {\bibfnamefont {K.~J.}\ \bibnamefont {Jiang}}, \bibinfo
  {author} {\bibfnamefont {L.}~\bibnamefont {Tarruell}}, \bibinfo {author}
  {\bibfnamefont {M.}~\bibnamefont {Teichmann}}, \bibinfo {author}
  {\bibfnamefont {J.}~\bibnamefont {McKeever}}, \bibinfo {author}
  {\bibfnamefont {F.}~\bibnamefont {Chevy}},\ and\ \bibinfo {author}
  {\bibfnamefont {C.}~\bibnamefont {Salomon}},\ }\bibfield  {title} {\bibinfo
  {title} {{Collective Oscillations of an Imbalanced Fermi Gas: Axial
  Compression Modes and Polaron Effective Mass}},\ }\href
  {https://doi.org/10.1103/PhysRevLett.103.170402} {\bibfield  {journal}
  {\bibinfo  {journal} {Phys. Rev. Lett.}\ }\textbf {\bibinfo {volume} {103}},\
  \bibinfo {pages} {170402} (\bibinfo {year} {2009})}\BibitemShut {NoStop}%
\bibitem [{\citenamefont {Hannaford}(2012)}]{Hannaford2012}%
  \BibitemOpen
  \bibfield  {author} {\bibinfo {author} {\bibfnamefont {P.}~\bibnamefont
  {Hannaford}},\ }\bibfield  {title} {\bibinfo {title} {{Repulsive polarons
  found}},\ }\href {https://doi.org/10.1038/nature11196} {\bibfield  {journal}
  {\bibinfo  {journal} {Nature}\ }\textbf {\bibinfo {volume} {485}},\ \bibinfo
  {pages} {588} (\bibinfo {year} {2012})}\BibitemShut {NoStop}%
\bibitem [{\citenamefont {Koschorreck}\ \emph {et~al.}(2012)\citenamefont
  {Koschorreck}, \citenamefont {Pertot}, \citenamefont {Vogt}, \citenamefont
  {Fr{\"o}hlich}, \citenamefont {Feld},\ and\ \citenamefont
  {K{\"o}hl}}]{Koschorreck2012}%
  \BibitemOpen
  \bibfield  {author} {\bibinfo {author} {\bibfnamefont {M.}~\bibnamefont
  {Koschorreck}}, \bibinfo {author} {\bibfnamefont {D.}~\bibnamefont {Pertot}},
  \bibinfo {author} {\bibfnamefont {E.}~\bibnamefont {Vogt}}, \bibinfo {author}
  {\bibfnamefont {B.}~\bibnamefont {Fr{\"o}hlich}}, \bibinfo {author}
  {\bibfnamefont {M.}~\bibnamefont {Feld}},\ and\ \bibinfo {author}
  {\bibfnamefont {M.}~\bibnamefont {K{\"o}hl}},\ }\bibfield  {title} {\bibinfo
  {title} {{Attractive and repulsive Fermi polarons in two dimensions}},\
  }\href {https://doi.org/10.1038/nature11151} {\bibfield  {journal} {\bibinfo
  {journal} {Nature}\ }\textbf {\bibinfo {volume} {485}},\ \bibinfo {pages}
  {619} (\bibinfo {year} {2012})}\BibitemShut {NoStop}%
\bibitem [{\citenamefont {Kohstall}\ \emph {et~al.}(2012)\citenamefont
  {Kohstall}, \citenamefont {Zaccanti}, \citenamefont {Jag}, \citenamefont
  {Trenkwalder}, \citenamefont {Massignan}, \citenamefont {Bruun},
  \citenamefont {Schreck},\ and\ \citenamefont {Grimm}}]{Kohstall2012}%
  \BibitemOpen
  \bibfield  {author} {\bibinfo {author} {\bibfnamefont {C.}~\bibnamefont
  {Kohstall}}, \bibinfo {author} {\bibfnamefont {M.}~\bibnamefont {Zaccanti}},
  \bibinfo {author} {\bibfnamefont {M.}~\bibnamefont {Jag}}, \bibinfo {author}
  {\bibfnamefont {A.}~\bibnamefont {Trenkwalder}}, \bibinfo {author}
  {\bibfnamefont {P.}~\bibnamefont {Massignan}}, \bibinfo {author}
  {\bibfnamefont {G.~M.}\ \bibnamefont {Bruun}}, \bibinfo {author}
  {\bibfnamefont {F.}~\bibnamefont {Schreck}},\ and\ \bibinfo {author}
  {\bibfnamefont {R.}~\bibnamefont {Grimm}},\ }\bibfield  {title} {\bibinfo
  {title} {{Metastability and coherence of repulsive polarons in a strongly
  interacting Fermi mixture}},\ }\href {https://doi.org/10.1038/nature11065}
  {\bibfield  {journal} {\bibinfo  {journal} {Nature}\ }\textbf {\bibinfo
  {volume} {485}},\ \bibinfo {pages} {615} (\bibinfo {year}
  {2012})}\BibitemShut {NoStop}%
\bibitem [{\citenamefont {Ness}\ \emph {et~al.}(2020)\citenamefont {Ness},
  \citenamefont {Shkedrov}, \citenamefont {Florshaim}, \citenamefont {Diessel},
  \citenamefont {von Milczewski}, \citenamefont {Schmidt},\ and\ \citenamefont
  {Sagi}}]{PhysRevX.10.041019}%
  \BibitemOpen
  \bibfield  {author} {\bibinfo {author} {\bibfnamefont {G.}~\bibnamefont
  {Ness}}, \bibinfo {author} {\bibfnamefont {C.}~\bibnamefont {Shkedrov}},
  \bibinfo {author} {\bibfnamefont {Y.}~\bibnamefont {Florshaim}}, \bibinfo
  {author} {\bibfnamefont {O.~K.}\ \bibnamefont {Diessel}}, \bibinfo {author}
  {\bibfnamefont {J.}~\bibnamefont {von Milczewski}}, \bibinfo {author}
  {\bibfnamefont {R.}~\bibnamefont {Schmidt}},\ and\ \bibinfo {author}
  {\bibfnamefont {Y.}~\bibnamefont {Sagi}},\ }\bibfield  {title} {\bibinfo
  {title} {{Observation of a Smooth Polaron-Molecule Transition in a Degenerate
  Fermi Gas}},\ }\href {https://doi.org/10.1103/PhysRevX.10.041019} {\bibfield
  {journal} {\bibinfo  {journal} {Phys. Rev. X}\ }\textbf {\bibinfo {volume}
  {10}},\ \bibinfo {pages} {041019} (\bibinfo {year} {2020})}\BibitemShut
  {NoStop}%
\bibitem [{\citenamefont {Adlong}\ \emph {et~al.}(2020)\citenamefont {Adlong},
  \citenamefont {Liu}, \citenamefont {Scazza}, \citenamefont {Zaccanti},
  \citenamefont {Oppong}, \citenamefont {F\"olling}, \citenamefont {Parish},\
  and\ \citenamefont {Levinsen}}]{PhysRevLett.125.133401}%
  \BibitemOpen
  \bibfield  {author} {\bibinfo {author} {\bibfnamefont {H.~S.}\ \bibnamefont
  {Adlong}}, \bibinfo {author} {\bibfnamefont {W.~E.}\ \bibnamefont {Liu}},
  \bibinfo {author} {\bibfnamefont {F.}~\bibnamefont {Scazza}}, \bibinfo
  {author} {\bibfnamefont {M.}~\bibnamefont {Zaccanti}}, \bibinfo {author}
  {\bibfnamefont {N.~D.}\ \bibnamefont {Oppong}}, \bibinfo {author}
  {\bibfnamefont {S.}~\bibnamefont {F\"olling}}, \bibinfo {author}
  {\bibfnamefont {M.~M.}\ \bibnamefont {Parish}},\ and\ \bibinfo {author}
  {\bibfnamefont {J.}~\bibnamefont {Levinsen}},\ }\bibfield  {title} {\bibinfo
  {title} {{Quasiparticle Lifetime of the Repulsive Fermi Polaron}},\ }\href
  {https://doi.org/10.1103/PhysRevLett.125.133401} {\bibfield  {journal}
  {\bibinfo  {journal} {Phys. Rev. Lett.}\ }\textbf {\bibinfo {volume} {125}},\
  \bibinfo {pages} {133401} (\bibinfo {year} {2020})}\BibitemShut {NoStop}%
\bibitem [{\citenamefont {Hu}\ and\ \citenamefont
  {Liu}(2022)}]{PhysRevA.106.063306}%
  \BibitemOpen
  \bibfield  {author} {\bibinfo {author} {\bibfnamefont {H.}~\bibnamefont
  {Hu}}\ and\ \bibinfo {author} {\bibfnamefont {X.-J.}\ \bibnamefont {Liu}},\
  }\bibfield  {title} {\bibinfo {title} {{Raman spectroscopy of Fermi
  polarons}},\ }\href {https://doi.org/10.1103/PhysRevA.106.063306} {\bibfield
  {journal} {\bibinfo  {journal} {Phys. Rev. A}\ }\textbf {\bibinfo {volume}
  {106}},\ \bibinfo {pages} {063306} (\bibinfo {year} {2022})}\BibitemShut
  {NoStop}%
\bibitem [{\citenamefont {Hu}\ \emph {et~al.}(2016)\citenamefont {Hu},
  \citenamefont {Van~de Graaff}, \citenamefont {Kedar}, \citenamefont {Corson},
  \citenamefont {Cornell},\ and\ \citenamefont {Jin}}]{PhysRevLett.117.055301}%
  \BibitemOpen
  \bibfield  {author} {\bibinfo {author} {\bibfnamefont {M.-G.}\ \bibnamefont
  {Hu}}, \bibinfo {author} {\bibfnamefont {M.~J.}\ \bibnamefont {Van~de
  Graaff}}, \bibinfo {author} {\bibfnamefont {D.}~\bibnamefont {Kedar}},
  \bibinfo {author} {\bibfnamefont {J.~P.}\ \bibnamefont {Corson}}, \bibinfo
  {author} {\bibfnamefont {E.~A.}\ \bibnamefont {Cornell}},\ and\ \bibinfo
  {author} {\bibfnamefont {D.~S.}\ \bibnamefont {Jin}},\ }\bibfield  {title}
  {\bibinfo {title} {{Bose Polarons in the Strongly Interacting Regime}},\
  }\href {https://doi.org/10.1103/PhysRevLett.117.055301} {\bibfield  {journal}
  {\bibinfo  {journal} {Phys. Rev. Lett.}\ }\textbf {\bibinfo {volume} {117}},\
  \bibinfo {pages} {055301} (\bibinfo {year} {2016})}\BibitemShut {NoStop}%
\bibitem [{\citenamefont {J\o{}rgensen}\ \emph {et~al.}(2016)\citenamefont
  {J\o{}rgensen}, \citenamefont {Wacker}, \citenamefont {Skalmstang},
  \citenamefont {Parish}, \citenamefont {Levinsen}, \citenamefont
  {Christensen}, \citenamefont {Bruun},\ and\ \citenamefont
  {Arlt}}]{PhysRevLett.117.055302}%
  \BibitemOpen
  \bibfield  {author} {\bibinfo {author} {\bibfnamefont {N.~B.}\ \bibnamefont
  {J\o{}rgensen}}, \bibinfo {author} {\bibfnamefont {L.}~\bibnamefont
  {Wacker}}, \bibinfo {author} {\bibfnamefont {K.~T.}\ \bibnamefont
  {Skalmstang}}, \bibinfo {author} {\bibfnamefont {M.~M.}\ \bibnamefont
  {Parish}}, \bibinfo {author} {\bibfnamefont {J.}~\bibnamefont {Levinsen}},
  \bibinfo {author} {\bibfnamefont {R.~S.}\ \bibnamefont {Christensen}},
  \bibinfo {author} {\bibfnamefont {G.~M.}\ \bibnamefont {Bruun}},\ and\
  \bibinfo {author} {\bibfnamefont {J.~J.}\ \bibnamefont {Arlt}},\ }\bibfield
  {title} {\bibinfo {title} {{Observation of Attractive and Repulsive Polarons
  in a Bose-Einstein Condensate}},\ }\href
  {https://doi.org/10.1103/PhysRevLett.117.055302} {\bibfield  {journal}
  {\bibinfo  {journal} {Phys. Rev. Lett.}\ }\textbf {\bibinfo {volume} {117}},\
  \bibinfo {pages} {055302} (\bibinfo {year} {2016})}\BibitemShut {NoStop}%
\bibitem [{\citenamefont {Pe\~na Ardila}\ \emph {et~al.}(2019)\citenamefont
  {Pe\~na Ardila}, \citenamefont {J\o{}rgensen}, \citenamefont {Pohl},
  \citenamefont {Giorgini}, \citenamefont {Bruun},\ and\ \citenamefont
  {Arlt}}]{PhysRevA.99.063607}%
  \BibitemOpen
  \bibfield  {author} {\bibinfo {author} {\bibfnamefont {L.~A.}\ \bibnamefont
  {Pe\~na Ardila}}, \bibinfo {author} {\bibfnamefont {N.~B.}\ \bibnamefont
  {J\o{}rgensen}}, \bibinfo {author} {\bibfnamefont {T.}~\bibnamefont {Pohl}},
  \bibinfo {author} {\bibfnamefont {S.}~\bibnamefont {Giorgini}}, \bibinfo
  {author} {\bibfnamefont {G.~M.}\ \bibnamefont {Bruun}},\ and\ \bibinfo
  {author} {\bibfnamefont {J.~J.}\ \bibnamefont {Arlt}},\ }\bibfield  {title}
  {\bibinfo {title} {{Analyzing a Bose polaron across resonant interactions}},\
  }\href {https://doi.org/10.1103/PhysRevA.99.063607} {\bibfield  {journal}
  {\bibinfo  {journal} {Phys. Rev. A}\ }\textbf {\bibinfo {volume} {99}},\
  \bibinfo {pages} {063607} (\bibinfo {year} {2019})}\BibitemShut {NoStop}%
\bibitem [{\citenamefont {Yan}\ \emph {et~al.}(2020)\citenamefont {Yan},
  \citenamefont {Ni}, \citenamefont {Robens},\ and\ \citenamefont
  {Zwierlein}}]{doi:10.1126/science.aax5850}%
  \BibitemOpen
  \bibfield  {author} {\bibinfo {author} {\bibfnamefont {Z.~Z.}\ \bibnamefont
  {Yan}}, \bibinfo {author} {\bibfnamefont {Y.}~\bibnamefont {Ni}}, \bibinfo
  {author} {\bibfnamefont {C.}~\bibnamefont {Robens}},\ and\ \bibinfo {author}
  {\bibfnamefont {M.~W.}\ \bibnamefont {Zwierlein}},\ }\bibfield  {title}
  {\bibinfo {title} {{Bose polarons near quantum criticality}},\ }\href
  {https://doi.org/10.1126/science.aax5850} {\bibfield  {journal} {\bibinfo
  {journal} {Science}\ }\textbf {\bibinfo {volume} {368}},\ \bibinfo {pages}
  {190} (\bibinfo {year} {2020})}\BibitemShut {NoStop}%
\bibitem [{\citenamefont {Skou}\ \emph {et~al.}(2022)\citenamefont {Skou},
  \citenamefont {Nielsen}, \citenamefont {Skov}, \citenamefont {Morgen},
  \citenamefont {J\o{}rgensen}, \citenamefont {Camacho-Guardian}, \citenamefont
  {Pohl}, \citenamefont {Bruun},\ and\ \citenamefont
  {Arlt}}]{PhysRevResearch.4.043093}%
  \BibitemOpen
  \bibfield  {author} {\bibinfo {author} {\bibfnamefont {M.~G.}\ \bibnamefont
  {Skou}}, \bibinfo {author} {\bibfnamefont {K.~K.}\ \bibnamefont {Nielsen}},
  \bibinfo {author} {\bibfnamefont {T.~G.}\ \bibnamefont {Skov}}, \bibinfo
  {author} {\bibfnamefont {A.~M.}\ \bibnamefont {Morgen}}, \bibinfo {author}
  {\bibfnamefont {N.~B.}\ \bibnamefont {J\o{}rgensen}}, \bibinfo {author}
  {\bibfnamefont {A.}~\bibnamefont {Camacho-Guardian}}, \bibinfo {author}
  {\bibfnamefont {T.}~\bibnamefont {Pohl}}, \bibinfo {author} {\bibfnamefont
  {G.~M.}\ \bibnamefont {Bruun}},\ and\ \bibinfo {author} {\bibfnamefont
  {J.~J.}\ \bibnamefont {Arlt}},\ }\bibfield  {title} {\bibinfo {title} {{Life
  and death of the Bose polaron}},\ }\href
  {https://doi.org/10.1103/PhysRevResearch.4.043093} {\bibfield  {journal}
  {\bibinfo  {journal} {Phys. Rev. Res.}\ }\textbf {\bibinfo {volume} {4}},\
  \bibinfo {pages} {043093} (\bibinfo {year} {2022})}\BibitemShut {NoStop}%
\bibitem [{\citenamefont {Yan}\ \emph {et~al.}(2024)\citenamefont {Yan},
  \citenamefont {Ni}, \citenamefont {Chuang}, \citenamefont {Dolgirev},
  \citenamefont {Seetharam}, \citenamefont {Demler}, \citenamefont {Robens},\
  and\ \citenamefont {Zwierlein}}]{Yan2024}%
  \BibitemOpen
  \bibfield  {author} {\bibinfo {author} {\bibfnamefont {Z.~Z.}\ \bibnamefont
  {Yan}}, \bibinfo {author} {\bibfnamefont {Y.}~\bibnamefont {Ni}}, \bibinfo
  {author} {\bibfnamefont {A.}~\bibnamefont {Chuang}}, \bibinfo {author}
  {\bibfnamefont {P.~E.}\ \bibnamefont {Dolgirev}}, \bibinfo {author}
  {\bibfnamefont {K.}~\bibnamefont {Seetharam}}, \bibinfo {author}
  {\bibfnamefont {E.}~\bibnamefont {Demler}}, \bibinfo {author} {\bibfnamefont
  {C.}~\bibnamefont {Robens}},\ and\ \bibinfo {author} {\bibfnamefont
  {M.}~\bibnamefont {Zwierlein}},\ }\bibfield  {title} {\bibinfo {title}
  {{Collective flow of fermionic impurities immersed in a Bose--Einstein
  condensate}},\ }\href {https://doi.org/10.1038/s41567-024-02541-w} {\bibfield
   {journal} {\bibinfo  {journal} {Nat. Phys.}\ }\textbf {\bibinfo {volume}
  {20}},\ \bibinfo {pages} {1395} (\bibinfo {year} {2024})}\BibitemShut
  {NoStop}%
\bibitem [{\citenamefont {Rath}\ and\ \citenamefont
  {Schmidt}(2013)}]{PhysRevA.88.053632}%
  \BibitemOpen
  \bibfield  {author} {\bibinfo {author} {\bibfnamefont {S.~P.}\ \bibnamefont
  {Rath}}\ and\ \bibinfo {author} {\bibfnamefont {R.}~\bibnamefont {Schmidt}},\
  }\bibfield  {title} {\bibinfo {title} {{Field-theoretical study of the Bose
  polaron}},\ }\href {https://doi.org/10.1103/PhysRevA.88.053632} {\bibfield
  {journal} {\bibinfo  {journal} {Phys. Rev. A}\ }\textbf {\bibinfo {volume}
  {88}},\ \bibinfo {pages} {053632} (\bibinfo {year} {2013})}\BibitemShut
  {NoStop}%
\bibitem [{\citenamefont {Li}\ and\ \citenamefont
  {Das~Sarma}(2014)}]{PhysRevA.90.013618}%
  \BibitemOpen
  \bibfield  {author} {\bibinfo {author} {\bibfnamefont {W.}~\bibnamefont
  {Li}}\ and\ \bibinfo {author} {\bibfnamefont {S.}~\bibnamefont {Das~Sarma}},\
  }\bibfield  {title} {\bibinfo {title} {{Variational study of polarons in
  Bose-Einstein condensates}},\ }\href
  {https://doi.org/10.1103/PhysRevA.90.013618} {\bibfield  {journal} {\bibinfo
  {journal} {Phys. Rev. A}\ }\textbf {\bibinfo {volume} {90}},\ \bibinfo
  {pages} {013618} (\bibinfo {year} {2014})}\BibitemShut {NoStop}%
\bibitem [{\citenamefont {Ardila}\ and\ \citenamefont
  {Giorgini}(2015)}]{PhysRevA.92.033612}%
  \BibitemOpen
  \bibfield  {author} {\bibinfo {author} {\bibfnamefont {L.~A. P.~n.}\
  \bibnamefont {Ardila}}\ and\ \bibinfo {author} {\bibfnamefont
  {S.}~\bibnamefont {Giorgini}},\ }\bibfield  {title} {\bibinfo {title}
  {{Impurity in a Bose-Einstein condensate: Study of the attractive and
  repulsive branch using quantum Monte Carlo methods}},\ }\href
  {https://doi.org/10.1103/PhysRevA.92.033612} {\bibfield  {journal} {\bibinfo
  {journal} {Phys. Rev. A}\ }\textbf {\bibinfo {volume} {92}},\ \bibinfo
  {pages} {033612} (\bibinfo {year} {2015})}\BibitemShut {NoStop}%
\bibitem [{\citenamefont {Drescher}\ \emph {et~al.}(2021)\citenamefont
  {Drescher}, \citenamefont {Salmhofer},\ and\ \citenamefont
  {Enss}}]{PhysRevA.103.033317}%
  \BibitemOpen
  \bibfield  {author} {\bibinfo {author} {\bibfnamefont {M.}~\bibnamefont
  {Drescher}}, \bibinfo {author} {\bibfnamefont {M.}~\bibnamefont
  {Salmhofer}},\ and\ \bibinfo {author} {\bibfnamefont {T.}~\bibnamefont
  {Enss}},\ }\bibfield  {title} {\bibinfo {title} {{Quench Dynamics of the
  Ideal Bose Polaron at Zero and Nonzero Temperatures}},\ }\href
  {https://doi.org/10.1103/PhysRevA.103.033317} {\bibfield  {journal} {\bibinfo
   {journal} {Phys. Rev. A}\ }\textbf {\bibinfo {volume} {103}},\ \bibinfo
  {pages} {033317} (\bibinfo {year} {2021})}\BibitemShut {NoStop}%
\bibitem [{\citenamefont {Isaule}\ \emph {et~al.}(2021)\citenamefont {Isaule},
  \citenamefont {Morera}, \citenamefont {Massignan},\ and\ \citenamefont
  {Juli\'a-D\'{\i}az}}]{PhysRevA.104.023317}%
  \BibitemOpen
  \bibfield  {author} {\bibinfo {author} {\bibfnamefont {F.}~\bibnamefont
  {Isaule}}, \bibinfo {author} {\bibfnamefont {I.}~\bibnamefont {Morera}},
  \bibinfo {author} {\bibfnamefont {P.}~\bibnamefont {Massignan}},\ and\
  \bibinfo {author} {\bibfnamefont {B.}~\bibnamefont {Juli\'a-D\'{\i}az}},\
  }\bibfield  {title} {\bibinfo {title} {{Renormalization-group study of Bose
  polarons}},\ }\href {https://doi.org/10.1103/PhysRevA.104.023317} {\bibfield
  {journal} {\bibinfo  {journal} {Phys. Rev. A}\ }\textbf {\bibinfo {volume}
  {104}},\ \bibinfo {pages} {023317} (\bibinfo {year} {2021})}\BibitemShut
  {NoStop}%
\bibitem [{\citenamefont {Ristivojevic}(2021)}]{PhysRevA.104.052218}%
  \BibitemOpen
  \bibfield  {author} {\bibinfo {author} {\bibfnamefont {Z.}~\bibnamefont
  {Ristivojevic}},\ }\bibfield  {title} {\bibinfo {title} {{Exact result for
  the polaron mass in a one-dimensional Bose gas}},\ }\href
  {https://doi.org/10.1103/PhysRevA.104.052218} {\bibfield  {journal} {\bibinfo
   {journal} {Phys. Rev. A}\ }\textbf {\bibinfo {volume} {104}},\ \bibinfo
  {pages} {052218} (\bibinfo {year} {2021})}\BibitemShut {NoStop}%
\bibitem [{\citenamefont {Pascual}\ and\ \citenamefont
  {Boronat}(2021)}]{PhysRevLett.127.205301}%
  \BibitemOpen
  \bibfield  {author} {\bibinfo {author} {\bibfnamefont {G.}~\bibnamefont
  {Pascual}}\ and\ \bibinfo {author} {\bibfnamefont {J.}~\bibnamefont
  {Boronat}},\ }\bibfield  {title} {\bibinfo {title} {{Quasiparticle Nature of
  the Bose Polaron at Finite Temperature}},\ }\href
  {https://doi.org/10.1103/PhysRevLett.127.205301} {\bibfield  {journal}
  {\bibinfo  {journal} {Phys. Rev. Lett.}\ }\textbf {\bibinfo {volume} {127}},\
  \bibinfo {pages} {205301} (\bibinfo {year} {2021})}\BibitemShut {NoStop}%
\bibitem [{\citenamefont {Yakaboylu}(2022)}]{PhysRevA.106.033321}%
  \BibitemOpen
  \bibfield  {author} {\bibinfo {author} {\bibfnamefont {E.}~\bibnamefont
  {Yakaboylu}},\ }\bibfield  {title} {\bibinfo {title} {{Analytical approach to
  the Bose polaron via a $q$-deformed Lie algebra}},\ }\href
  {https://doi.org/10.1103/PhysRevA.106.033321} {\bibfield  {journal} {\bibinfo
   {journal} {Phys. Rev. A}\ }\textbf {\bibinfo {volume} {106}},\ \bibinfo
  {pages} {033321} (\bibinfo {year} {2022})}\BibitemShut {NoStop}%
\bibitem [{\citenamefont {Nakano}\ \emph {et~al.}(2024)\citenamefont {Nakano},
  \citenamefont {Parish},\ and\ \citenamefont
  {Levinsen}}]{PhysRevA.109.013325}%
  \BibitemOpen
  \bibfield  {author} {\bibinfo {author} {\bibfnamefont {Y.}~\bibnamefont
  {Nakano}}, \bibinfo {author} {\bibfnamefont {M.~M.}\ \bibnamefont {Parish}},\
  and\ \bibinfo {author} {\bibfnamefont {J.}~\bibnamefont {Levinsen}},\
  }\bibfield  {title} {\bibinfo {title} {{Variational approach to the
  two-dimensional Bose polaron}},\ }\href
  {https://doi.org/10.1103/PhysRevA.109.013325} {\bibfield  {journal} {\bibinfo
   {journal} {Phys. Rev. A}\ }\textbf {\bibinfo {volume} {109}},\ \bibinfo
  {pages} {013325} (\bibinfo {year} {2024})}\BibitemShut {NoStop}%
\bibitem [{\citenamefont {Yegovtsev}\ \emph {et~al.}(2024)\citenamefont
  {Yegovtsev}, \citenamefont {Astrakharchik}, \citenamefont {Massignan},\ and\
  \citenamefont {Gurarie}}]{PhysRevA.110.023310}%
  \BibitemOpen
  \bibfield  {author} {\bibinfo {author} {\bibfnamefont {N.}~\bibnamefont
  {Yegovtsev}}, \bibinfo {author} {\bibfnamefont {G.~E.}\ \bibnamefont
  {Astrakharchik}}, \bibinfo {author} {\bibfnamefont {P.}~\bibnamefont
  {Massignan}},\ and\ \bibinfo {author} {\bibfnamefont {V.}~\bibnamefont
  {Gurarie}},\ }\bibfield  {title} {\bibinfo {title} {{Exact results for heavy
  unitary Bose polarons}},\ }\href
  {https://doi.org/10.1103/PhysRevA.110.023310} {\bibfield  {journal} {\bibinfo
   {journal} {Phys. Rev. A}\ }\textbf {\bibinfo {volume} {110}},\ \bibinfo
  {pages} {023310} (\bibinfo {year} {2024})}\BibitemShut {NoStop}%
\bibitem [{\citenamefont {Dehkharghani}\ \emph {et~al.}(2018)\citenamefont
  {Dehkharghani}, \citenamefont {Volosniev},\ and\ \citenamefont
  {Zinner}}]{PhysRevLett.121.080405}%
  \BibitemOpen
  \bibfield  {author} {\bibinfo {author} {\bibfnamefont {A.~S.}\ \bibnamefont
  {Dehkharghani}}, \bibinfo {author} {\bibfnamefont {A.~G.}\ \bibnamefont
  {Volosniev}},\ and\ \bibinfo {author} {\bibfnamefont {N.~T.}\ \bibnamefont
  {Zinner}},\ }\bibfield  {title} {\bibinfo {title} {{Coalescence of Two
  Impurities in a Trapped One-dimensional Bose Gas}},\ }\href
  {https://doi.org/10.1103/PhysRevLett.121.080405} {\bibfield  {journal}
  {\bibinfo  {journal} {Phys. Rev. Lett.}\ }\textbf {\bibinfo {volume} {121}},\
  \bibinfo {pages} {080405} (\bibinfo {year} {2018})}\BibitemShut {NoStop}%
\bibitem [{\citenamefont {Camacho-Guardian}\ \emph {et~al.}(2018)\citenamefont
  {Camacho-Guardian}, \citenamefont {Pe\~na Ardila}, \citenamefont {Pohl},\
  and\ \citenamefont {Bruun}}]{PhysRevLett.121.013401}%
  \BibitemOpen
  \bibfield  {author} {\bibinfo {author} {\bibfnamefont {A.}~\bibnamefont
  {Camacho-Guardian}}, \bibinfo {author} {\bibfnamefont {L.~A.}\ \bibnamefont
  {Pe\~na Ardila}}, \bibinfo {author} {\bibfnamefont {T.}~\bibnamefont
  {Pohl}},\ and\ \bibinfo {author} {\bibfnamefont {G.~M.}\ \bibnamefont
  {Bruun}},\ }\bibfield  {title} {\bibinfo {title} {{Bipolarons in a
  Bose-Einstein Condensate}},\ }\href
  {https://doi.org/10.1103/PhysRevLett.121.013401} {\bibfield  {journal}
  {\bibinfo  {journal} {Phys. Rev. Lett.}\ }\textbf {\bibinfo {volume} {121}},\
  \bibinfo {pages} {013401} (\bibinfo {year} {2018})}\BibitemShut {NoStop}%
\bibitem [{\citenamefont {Camacho-Guardian}\ and\ \citenamefont
  {Bruun}(2018)}]{PhysRevX.8.031042}%
  \BibitemOpen
  \bibfield  {author} {\bibinfo {author} {\bibfnamefont {A.}~\bibnamefont
  {Camacho-Guardian}}\ and\ \bibinfo {author} {\bibfnamefont {G.~M.}\
  \bibnamefont {Bruun}},\ }\bibfield  {title} {\bibinfo {title} {{Landau
  Effective Interaction between Quasiparticles in a Bose-Einstein
  Condensate}},\ }\href {https://doi.org/10.1103/PhysRevX.8.031042} {\bibfield
  {journal} {\bibinfo  {journal} {Phys. Rev. X}\ }\textbf {\bibinfo {volume}
  {8}},\ \bibinfo {pages} {031042} (\bibinfo {year} {2018})}\BibitemShut
  {NoStop}%
\bibitem [{\citenamefont {Naidon}(2018)}]{doi:10.7566/JPSJ.87.043002}%
  \BibitemOpen
  \bibfield  {author} {\bibinfo {author} {\bibfnamefont {P.}~\bibnamefont
  {Naidon}},\ }\bibfield  {title} {\bibinfo {title} {{Two Impurities in a
  Bose–Einstein Condensate: From Yukawa to Efimov Attracted Polarons}},\
  }\href {https://doi.org/10.7566/JPSJ.87.043002} {\bibfield  {journal}
  {\bibinfo  {journal} {J. Phys. Soc. Jpn.}\ }\textbf {\bibinfo {volume}
  {87}},\ \bibinfo {pages} {043002} (\bibinfo {year} {2018})}\BibitemShut
  {NoStop}%
\bibitem [{\citenamefont {Jager}\ and\ \citenamefont
  {Barnett}(2022)}]{Jager_2022}%
  \BibitemOpen
  \bibfield  {author} {\bibinfo {author} {\bibfnamefont {J.}~\bibnamefont
  {Jager}}\ and\ \bibinfo {author} {\bibfnamefont {R.}~\bibnamefont
  {Barnett}},\ }\bibfield  {title} {\bibinfo {title} {{The effect of
  boson–boson interaction on the bipolaron formation}},\ }\href
  {https://doi.org/10.1088/1367-2630/ac9804} {\bibfield  {journal} {\bibinfo
  {journal} {New J. Phys.}\ }\textbf {\bibinfo {volume} {24}},\ \bibinfo
  {pages} {103032} (\bibinfo {year} {2022})}\BibitemShut {NoStop}%
\bibitem [{\citenamefont {Drescher}\ \emph {et~al.}(2023)\citenamefont
  {Drescher}, \citenamefont {Salmhofer},\ and\ \citenamefont
  {Enss}}]{PhysRevA.107.063301}%
  \BibitemOpen
  \bibfield  {author} {\bibinfo {author} {\bibfnamefont {M.}~\bibnamefont
  {Drescher}}, \bibinfo {author} {\bibfnamefont {M.}~\bibnamefont
  {Salmhofer}},\ and\ \bibinfo {author} {\bibfnamefont {T.}~\bibnamefont
  {Enss}},\ }\bibfield  {title} {\bibinfo {title} {{Medium-induced interaction
  between impurities in a Bose-Einstein condensate}},\ }\href
  {https://doi.org/10.1103/PhysRevA.107.063301} {\bibfield  {journal} {\bibinfo
   {journal} {Phys. Rev. A}\ }\textbf {\bibinfo {volume} {107}},\ \bibinfo
  {pages} {063301} (\bibinfo {year} {2023})}\BibitemShut {NoStop}%
\bibitem [{\citenamefont {Yegovtsev}\ and\ \citenamefont
  {Gurarie}(2023)}]{PhysRevA.108.L051301}%
  \BibitemOpen
  \bibfield  {author} {\bibinfo {author} {\bibfnamefont {N.}~\bibnamefont
  {Yegovtsev}}\ and\ \bibinfo {author} {\bibfnamefont {V.}~\bibnamefont
  {Gurarie}},\ }\bibfield  {title} {\bibinfo {title} {{Effective mass and
  interaction energy of heavy Bose polarons at unitarity}},\ }\href
  {https://doi.org/10.1103/PhysRevA.108.L051301} {\bibfield  {journal}
  {\bibinfo  {journal} {Phys. Rev. A}\ }\textbf {\bibinfo {volume} {108}},\
  \bibinfo {pages} {L051301} (\bibinfo {year} {2023})}\BibitemShut {NoStop}%
\bibitem [{\citenamefont {Fujii}\ \emph {et~al.}(2022)\citenamefont {Fujii},
  \citenamefont {Hongo},\ and\ \citenamefont {Enss}}]{PhysRevLett.129.233401}%
  \BibitemOpen
  \bibfield  {author} {\bibinfo {author} {\bibfnamefont {K.}~\bibnamefont
  {Fujii}}, \bibinfo {author} {\bibfnamefont {M.}~\bibnamefont {Hongo}},\ and\
  \bibinfo {author} {\bibfnamefont {T.}~\bibnamefont {Enss}},\ }\bibfield
  {title} {\bibinfo {title} {{Universal van der Waals Force between Heavy
  Polarons in Superfluids}},\ }\href
  {https://doi.org/10.1103/PhysRevLett.129.233401} {\bibfield  {journal}
  {\bibinfo  {journal} {Phys. Rev. Lett.}\ }\textbf {\bibinfo {volume} {129}},\
  \bibinfo {pages} {233401} (\bibinfo {year} {2022})}\BibitemShut {NoStop}%
\bibitem [{\citenamefont {Paredes}\ \emph {et~al.}(2024)\citenamefont
  {Paredes}, \citenamefont {Bruun},\ and\ \citenamefont
  {Camacho-Guardian}}]{PhysRevA.110.030101}%
  \BibitemOpen
  \bibfield  {author} {\bibinfo {author} {\bibfnamefont {R.}~\bibnamefont
  {Paredes}}, \bibinfo {author} {\bibfnamefont {G.}~\bibnamefont {Bruun}},\
  and\ \bibinfo {author} {\bibfnamefont {A.}~\bibnamefont {Camacho-Guardian}},\
  }\bibfield  {title} {\bibinfo {title} {{Interactions mediated by atoms,
  photons, electrons, and excitons}},\ }\href
  {https://doi.org/10.1103/PhysRevA.110.030101} {\bibfield  {journal} {\bibinfo
   {journal} {Phys. Rev. A}\ }\textbf {\bibinfo {volume} {110}},\ \bibinfo
  {pages} {030101} (\bibinfo {year} {2024})}\BibitemShut {NoStop}%
\bibitem [{\citenamefont {Panochko}\ and\ \citenamefont
  {Pastukhov}(2021)}]{Panochko_2021}%
  \BibitemOpen
  \bibfield  {author} {\bibinfo {author} {\bibfnamefont {G.}~\bibnamefont
  {Panochko}}\ and\ \bibinfo {author} {\bibfnamefont {V.}~\bibnamefont
  {Pastukhov}},\ }\bibfield  {title} {\bibinfo {title} {{Two- and three-body
  effective potentials between impurities in ideal BEC}},\ }\href
  {https://doi.org/10.1088/1751-8121/abdbc5} {\bibfield  {journal} {\bibinfo
  {journal} {J. Phys. A: Math. Theor.}\ }\textbf {\bibinfo {volume} {54}},\
  \bibinfo {pages} {085001} (\bibinfo {year} {2021})}\BibitemShut {NoStop}%
\bibitem [{\citenamefont {Panochko}\ and\ \citenamefont
  {Pastukhov}(2022)}]{atoms10010019}%
  \BibitemOpen
  \bibfield  {author} {\bibinfo {author} {\bibfnamefont {G.}~\bibnamefont
  {Panochko}}\ and\ \bibinfo {author} {\bibfnamefont {V.}~\bibnamefont
  {Pastukhov}},\ }\bibfield  {title} {\bibinfo {title} {{Static Impurities in a
  Weakly Interacting Bose Gas}},\ }\href
  {https://www.mdpi.com/2218-2004/10/1/19} {\bibfield  {journal} {\bibinfo
  {journal} {Atoms}\ }\textbf {\bibinfo {volume} {10}},\ \bibinfo {pages} {19}
  (\bibinfo {year} {2022})}\BibitemShut {NoStop}%
\bibitem [{\citenamefont {Chen}\ \emph {et~al.}(2023)\citenamefont {Chen},
  \citenamefont {Li}, \citenamefont {Sun}, \citenamefont {Chen}, \citenamefont
  {Chang},\ and\ \citenamefont {Tung}}]{PhysRevA.108.033301}%
  \BibitemOpen
  \bibfield  {author} {\bibinfo {author} {\bibfnamefont {Y.-D.}\ \bibnamefont
  {Chen}}, \bibinfo {author} {\bibfnamefont {W.-X.}\ \bibnamefont {Li}},
  \bibinfo {author} {\bibfnamefont {Y.-T.}\ \bibnamefont {Sun}}, \bibinfo
  {author} {\bibfnamefont {Q.-C.}\ \bibnamefont {Chen}}, \bibinfo {author}
  {\bibfnamefont {P.-Y.}\ \bibnamefont {Chang}},\ and\ \bibinfo {author}
  {\bibfnamefont {S.}~\bibnamefont {Tung}},\ }\bibfield  {title} {\bibinfo
  {title} {{Dual-species Bose-Einstein condensates of $^{7}\mathrm{Li}$ and
  $^{133}\mathrm{Cs}$}},\ }\href {https://doi.org/10.1103/PhysRevA.108.033301}
  {\bibfield  {journal} {\bibinfo  {journal} {Phys. Rev. A}\ }\textbf {\bibinfo
  {volume} {108}},\ \bibinfo {pages} {033301} (\bibinfo {year}
  {2023})}\BibitemShut {NoStop}%
\bibitem [{\citenamefont {Hugenholtz}\ and\ \citenamefont
  {Pines}(1959)}]{PhysRev.116.489}%
  \BibitemOpen
  \bibfield  {author} {\bibinfo {author} {\bibfnamefont {N.~M.}\ \bibnamefont
  {Hugenholtz}}\ and\ \bibinfo {author} {\bibfnamefont {D.}~\bibnamefont
  {Pines}},\ }\bibfield  {title} {\bibinfo {title} {Ground-state energy and
  excitation spectrum of a system of interacting bosons},\ }\href
  {https://doi.org/10.1103/PhysRev.116.489} {\bibfield  {journal} {\bibinfo
  {journal} {Phys. Rev.}\ }\textbf {\bibinfo {volume} {116}},\ \bibinfo {pages}
  {489} (\bibinfo {year} {1959})}\BibitemShut {NoStop}%
\bibitem [{sup()}]{supp}%
  \BibitemOpen
  \href@noop {} {}\bibinfo {note} {See Supplemental Material for the details of
  the derivation for the self-energy using Wilsonian renormalization, and the
  analysis of the self-energy, and for the derivation of various effective
  potentials, which includes Refs.~\cite{PhysRevA.88.053632, Panochko_2021,
  PhysRevA.107.063301, PhysRevA.79.013629, PhysRevA.102.063321,
  PhysRevA.107.063301, doi:10.7566/JPSJ.87.043002, atoms10010019,
  PhysRevX.8.031042, doi:10.7566/JPSJ.87.043002, Jager_2022,
  PhysRevA.107.063301}}\BibitemShut {NoStop}%
\bibitem [{\citenamefont {Nishida}(2009)}]{PhysRevA.79.013629}%
  \BibitemOpen
  \bibfield  {author} {\bibinfo {author} {\bibfnamefont {Y.}~\bibnamefont
  {Nishida}},\ }\bibfield  {title} {\bibinfo {title} {{Casimir interaction
  among heavy fermions in the BCS-BEC crossover}},\ }\href
  {https://doi.org/10.1103/PhysRevA.79.013629} {\bibfield  {journal} {\bibinfo
  {journal} {Phys. Rev. A}\ }\textbf {\bibinfo {volume} {79}},\ \bibinfo
  {pages} {013629} (\bibinfo {year} {2009})}\BibitemShut {NoStop}%
\bibitem [{\citenamefont {Enss}\ \emph {et~al.}(2020)\citenamefont {Enss},
  \citenamefont {Tran}, \citenamefont {Rautenberg}, \citenamefont {Gerken},
  \citenamefont {Lippi}, \citenamefont {Drescher}, \citenamefont {Zhu},
  \citenamefont {Weidem\"uller},\ and\ \citenamefont
  {Salmhofer}}]{PhysRevA.102.063321}%
  \BibitemOpen
  \bibfield  {author} {\bibinfo {author} {\bibfnamefont {T.}~\bibnamefont
  {Enss}}, \bibinfo {author} {\bibfnamefont {B.}~\bibnamefont {Tran}}, \bibinfo
  {author} {\bibfnamefont {M.}~\bibnamefont {Rautenberg}}, \bibinfo {author}
  {\bibfnamefont {M.}~\bibnamefont {Gerken}}, \bibinfo {author} {\bibfnamefont
  {E.}~\bibnamefont {Lippi}}, \bibinfo {author} {\bibfnamefont
  {M.}~\bibnamefont {Drescher}}, \bibinfo {author} {\bibfnamefont
  {B.}~\bibnamefont {Zhu}}, \bibinfo {author} {\bibfnamefont {M.}~\bibnamefont
  {Weidem\"uller}},\ and\ \bibinfo {author} {\bibfnamefont {M.}~\bibnamefont
  {Salmhofer}},\ }\bibfield  {title} {\bibinfo {title} {{Scattering of two
  heavy Fermi polarons: Resonances and quasibound states}},\ }\href
  {https://doi.org/10.1103/PhysRevA.102.063321} {\bibfield  {journal} {\bibinfo
   {journal} {Phys. Rev. A}\ }\textbf {\bibinfo {volume} {102}},\ \bibinfo
  {pages} {063321} (\bibinfo {year} {2020})}\BibitemShut {NoStop}%
\bibitem [{\citenamefont {Levinsen}\ \emph {et~al.}(2021)\citenamefont
  {Levinsen}, \citenamefont {Ardila}, \citenamefont {Yoshida},\ and\
  \citenamefont {Parish}}]{PhysRevLett.127.033401}%
  \BibitemOpen
  \bibfield  {author} {\bibinfo {author} {\bibfnamefont {J.}~\bibnamefont
  {Levinsen}}, \bibinfo {author} {\bibfnamefont {L.~A. P.~n.}\ \bibnamefont
  {Ardila}}, \bibinfo {author} {\bibfnamefont {S.~M.}\ \bibnamefont
  {Yoshida}},\ and\ \bibinfo {author} {\bibfnamefont {M.~M.}\ \bibnamefont
  {Parish}},\ }\bibfield  {title} {\bibinfo {title} {{Quantum Behavior of a
  Heavy Impurity Strongly Coupled to a Bose Gas}},\ }\href
  {https://doi.org/10.1103/PhysRevLett.127.033401} {\bibfield  {journal}
  {\bibinfo  {journal} {Phys. Rev. Lett.}\ }\textbf {\bibinfo {volume} {127}},\
  \bibinfo {pages} {033401} (\bibinfo {year} {2021})}\BibitemShut {NoStop}%
\bibitem [{\citenamefont {Shi}\ \emph {et~al.}(2018)\citenamefont {Shi},
  \citenamefont {Yoshida}, \citenamefont {Parish},\ and\ \citenamefont
  {Levinsen}}]{PhysRevLett.121.243401}%
  \BibitemOpen
  \bibfield  {author} {\bibinfo {author} {\bibfnamefont {Z.-Y.}\ \bibnamefont
  {Shi}}, \bibinfo {author} {\bibfnamefont {S.~M.}\ \bibnamefont {Yoshida}},
  \bibinfo {author} {\bibfnamefont {M.~M.}\ \bibnamefont {Parish}},\ and\
  \bibinfo {author} {\bibfnamefont {J.}~\bibnamefont {Levinsen}},\ }\bibfield
  {title} {\bibinfo {title} {{Impurity-Induced Multibody Resonances in a Bose
  Gas}},\ }\href {https://doi.org/10.1103/PhysRevLett.121.243401} {\bibfield
  {journal} {\bibinfo  {journal} {Phys. Rev. Lett.}\ }\textbf {\bibinfo
  {volume} {121}},\ \bibinfo {pages} {243401} (\bibinfo {year}
  {2018})}\BibitemShut {NoStop}%
\bibitem [{\citenamefont {Yoshida}\ \emph {et~al.}(2018)\citenamefont
  {Yoshida}, \citenamefont {Shi}, \citenamefont {Levinsen},\ and\ \citenamefont
  {Parish}}]{PhysRevA.98.062705}%
  \BibitemOpen
  \bibfield  {author} {\bibinfo {author} {\bibfnamefont {S.~M.}\ \bibnamefont
  {Yoshida}}, \bibinfo {author} {\bibfnamefont {Z.-Y.}\ \bibnamefont {Shi}},
  \bibinfo {author} {\bibfnamefont {J.}~\bibnamefont {Levinsen}},\ and\
  \bibinfo {author} {\bibfnamefont {M.~M.}\ \bibnamefont {Parish}},\ }\bibfield
   {title} {\bibinfo {title} {{Few-body states of bosons interacting with a
  heavy quantum impurity}},\ }\href
  {https://doi.org/10.1103/PhysRevA.98.062705} {\bibfield  {journal} {\bibinfo
  {journal} {Phys. Rev. A}\ }\textbf {\bibinfo {volume} {98}},\ \bibinfo
  {pages} {062705} (\bibinfo {year} {2018})}\BibitemShut {NoStop}%
\bibitem [{\citenamefont {Kaufman}\ and\ \citenamefont
  {Ni}(2021)}]{Kaufman2021}%
  \BibitemOpen
  \bibfield  {author} {\bibinfo {author} {\bibfnamefont {A.~M.}\ \bibnamefont
  {Kaufman}}\ and\ \bibinfo {author} {\bibfnamefont {K.-K.}\ \bibnamefont
  {Ni}},\ }\bibfield  {title} {\bibinfo {title} {{Quantum science with optical
  tweezer arrays of ultracold atoms and molecules}},\ }\href
  {https://doi.org/10.1038/s41567-021-01357-2} {\bibfield  {journal} {\bibinfo
  {journal} {Nat. Phys.}\ }\textbf {\bibinfo {volume} {17}},\ \bibinfo {pages}
  {1324} (\bibinfo {year} {2021})}\BibitemShut {NoStop}%
\end{thebibliography}
%


\clearpage
\appendix
\onecolumngrid
\setcounter{equation}{0}
\setcounter{figure}{0}
\setcounter{table}{0}
\setcounter{page}{1}

\renewcommand{\theequation}{S\arabic{equation}}
\renewcommand{\thefigure}{S\arabic{figure}}
\renewcommand{\thetable}{S\arabic{table}}

\begin{center}
  \textbf{\large Supplemental Materials:\\
  Asymptotic Freedom of Two Heavy Impurities in a Bose-Einstein
  Condensate}
\end{center}
\begin{center}
  Dong-Chen Zheng,$^{1,2}$ 
  Lin Wen,$^{3}$  and
  Renyuan Liao$^{1,2}$ \\
  \emph{\small $^1$Fujian Provincial Key Laboratory for Quantum Manipulation and New Energy Materials, College of Physics and Energy, Fujian Normal University, Fuzhou 350117, China}\\
  \emph{\small $^2$Fujian Provincial Collaborative Innovation Center for Advanced High-Field Superconducting Materials and Engineering, Fuzhou, 350117, China}\\
  \emph{\small $^3$College of Physics and Electronic Engineering, Chongqing Normal University, Chongqing 401331, China}
\end{center}

The effective theory with a momentum cutoff $\Lambda$ is given by
\begin{subequations}
\begin{align}
  & S_{\Lambda} = \frac{1}{2}
    \hspace{-0.8em}
    \sum_{\substack{\omega_{n} \\ 0<|\bp|\leqslant\Lambda}}
  \hspace{-0.8em}
  \Phi_{p}^{\dag}[-G_{B}^{-1}(p)]\Phi_{p}
  + \beta\Sigma_{I}(\Lambda) + \frac{g_{I}(\Lambda)}{V}
  \hspace{-1.5em}
  \sum_{\substack{\omega_{n} \\ |\bp_{1}|,|\bp_{2}|\leqslant\Lambda}}
  \hspace{-1em}
  (1-\delta_{\bp_{1}}\delta_{\bp_{2}})    
  U(\bp_{2}-\bp_{1})\aphi_{p_{1}}\phi_{p_{2}}, \\
  & G_{B}^{-1}(p) =
    \begin{pmatrix}
      i\omega_{n}-\frac{\bp^{2}}{2m_{B}}-g_{B}n_{B} & -g_{B}n_{B} \\
      -g_{B}n_{B} & -i\omega_{n}-\frac{\bp^{2}}{2m_{B}}-g_{B}n_{B}
    \end{pmatrix},
  \hspace{3em}\Phi_{p} =
    \begin{pmatrix}
      \phi_{p} \\
      \aphi_{-p}
    \end{pmatrix}, \\
  & U(\bp_{2}-\bp_{1}) =
    e^{i(\bp_{1}-\bp_{2})\cdot\br_{1}} +
    e^{i(\bp_{1}-\bp_{2})\cdot\br_{2}},
\end{align}
\label{eq:SM_S_Lambda}
\end{subequations}
where $p\equiv(\bp,i\omega_{n})$.
Note that the frequecy indices are the same in each term, i.e.,
$\aphi_{p_{1}}\phi_{p_{2}} =
\aphi(\bp_{1},i\omega_{n})\phi(\bp_{2},i\omega_{n})$.
Here $\Sigma_{I}(\Lambda)$ and $g_{I}(\Lambda)$ are the parameters
that we are going to renormalize in the theory.

Next, we impose a hard cutoff $\Lambda'=\Lambda - \Delta\Lambda$ to
the low-energy theory by decomposing the bosonic fields as
\begin{subequations}
\begin{align}
  & \phi_{p} = \phis_{p} + \phif_{p},
    \hspace{2em} 0<|\bp|\leqslant\Lambda, \\
  & \phis_{p} = \phi_{p}\Theta(\Lambda'-|\bp|), \\
  & \phif_{p} = \phi_{p}\Theta(|\bp|-\Lambda'),   
\end{align}  
\end{subequations}
where $\Theta(x)$ is the Heaviside step function.
We then divide the action into three parts,

\begin{subequations}
\begin{align}
  & S_{\Lambda} = S_{\s} + S_{\f} + S_{I},  \\
  & S_{\s} = \frac{1}{2}
  \hspace{-0.8em}
  \sum_{\substack{\omega_{n} \\ 0<|\bp|\leqslant\Lambda'}}
  \hspace{-0.8em}
  \Phi_{p}^{(\mathrm{s})\dag}[-G_{B}^{-1}(p)]\Phi_{p}^{(\mathrm{s})}
  + \beta\Sigma_{I}(\Lambda)
  + \frac{g_{I}(\Lambda)}{V}
    \hspace{-1.5em}
    \sum_{\substack{\omega_{n} \\ |\bp_{1}|,|\bp_{2}|\leqslant\Lambda'}}
    \hspace{-1em}
    (1-\delta_{\bp_{1}}\delta_{\bp_{2}})    
  U(\bp_{2}-\bp_{1})\aphis_{p_{1}}\phis_{p_{2}}, \\
  & S_{\f} = \frac{1}{2}
  \hspace{-0.8em}
  \sum_{\substack{\omega_{n} \\ \Lambda'<|\bp|\leqslant\Lambda}}
  \hspace{-0.8em}
  \Phi_{p}^{(\mathrm{f})\dag}[-G_{B}^{-1}(p)]\Phi_{p}^{(\mathrm{f})},\\
  & S_{I} = \sum_{p_{1},p_{2}} \frac{g_{I}(\Lambda)}{V}
    U(\bp_{2}-\bp_{1})\left(
    \aphis_{p_{1}}\phif_{p_{2}} + \aphif_{p_{1}}\phis_{p_{2}} +
    \aphif_{p_{1}}\phif_{p_{2}} \right)
    = \sum_{p_{1},p_{2}} \cL_{I}(p_{1},p_{2}),
\end{align}
\end{subequations}
where $\Phi^{(\mathrm{s})}_{p} \equiv (\phis_{p},\aphis_{-p})^{T}$ and
$\Phi^{(\mathrm{f})}_{p} \equiv (\phif_{p},\aphif_{-p})^{T}$.
\begin{table}[h]
\caption{Definitions of Diagrams}
\begin{tabular}{|c|c|c|c|}
  \hline Bosonic Fields
  & Creation & Annihilation
  \\ \hline
  Slow Mode
  & $\DefLegLong{(v0)}{fs}{edge label=\(p\)}{(v)} \aphis_{p}$
  & $\DefLegLong{(v)}{fs}{edge label'=\(p\)}{(v0)} \phis_{p}$
  \\ \hline
  Fast Mode
  & $\DefLegLong{(v0)}{ff}{edge label=\(p\)}{(v)} \aphif_{p}$
  & $\DefLegLong{(v)}{ff}{edge label'=\(p\)}{(v0)} \phif_{p}$
  \\ \hline
  $\frac{g_{I}}{V}U(\bp_{2}-\bp_{1})\delta_{\omega_{p_{2}}-\omega_{p_{1}}}$
  & \multicolumn{2}{c|}{\parbox[c][1cm][b]{2cm}{
    $\DefImpBoseInt{fs}{\(p_{2}\)}{fs}{\(p_{1}\)}$
    }}
  \\ \hline
\end{tabular}
\label{tab:DEF_FD}
\end{table}
For simplicity, we introduce Feynman diagrams, as defined in
TABLE~\ref{tab:DEF_FD}, to perform the calculations.
Using the diagrams, $\cL_{I}(p_{1},p_{2})$ can be cast as
\begin{equation}
  \mathcal{L}_{I}(p_{1},p_{2}) =
  \LI{ff}{fs} + \LI{fs}{ff} + \LI{ff}{ff}.
\end{equation}

We are looking for the effective action $S_{\Lambda'}$ with new cutoff
$\Lambda' = \Lambda-\Delta\Lambda$ based on 
\begin{align}
  \mathcal{Z}
  & = \int \cD[\aphi,\phi] e^{-S_{\Lambda}}
    = \mathcal{N} \int \cD[\aphis,\phis] e^{-S_{\s}}
    \frac{\int \cD[\aphif,\phif] e^{-S_{\f}}e^{-S_{I}}}
    {\int \cD[\aphif,\phif] e^{-S_{\f}}}
    = \mathcal{N} \int \cD[\aphis,\phis] e^{-S_{\s}}
    \langle e^{-S_{I}} \rangle_{\f} \nonumber \\
  & = \mathcal{N} \int \cD[\aphis,\phis] e^{-S_{\Lambda'}},
\end{align}
where $\mathcal{N}$ is an irrelevant constant, and the fast-mode
average of $X$ is defined by
\begin{align}
  \langle X \rangle_{\f} \equiv
  \frac{\int\cD[\aphif,\phif]e^{-S_{\f}} X}
  {\int\cD[\aphif,\phif]e^{-S_{\f}}}.
\end{align}
Applying the cumulant expansion, then the effective action
$S_{\Lambda'}$ can be approximated as
\begin{align}
  S_{\Lambda'} \approx S_{\s} + \langle S_{I} \rangle_{\f}
  - \frac{1}{2} \left(
  \langle S_{I}^{2} \rangle_{\f} - \langle S_{I} \rangle_{\f}^{2}
  \right). \label{eq:SM_S_Lambda_Prime_Form}
\end{align}

\section{Feynman Rules}

\newcommand{\acJ}{\mathcal{J}^{\ast}}
\newcommand{\cJ}{\mathcal{J}}

To investigate the fast-mode average, it is convenient to introduce
the fast-mode generating function
\begin{align}
  \mathcal{Z}_{\f}[\acJ,\cJ]
  \equiv &
  \int\cD[\aphif,\phif] \exp\left[
  \frac{1}{2}
  \sum_{p}
  \Phi_{p}^{(\mathrm{f})\dag}G_{B}^{-1}(p)\Phi_{p}^{(\mathrm{f})}
  +\sum_{p}\left(\acJ_{p}\phif_{p}+\aphif_{p}\cJ_{p}\right)
  \right] \nonumber \\
  = &
  \int\cD[\aphif,\phif] \exp\left[
  \frac{1}{2}
  \sum_{p} \left(
  \Phi_{p}^{(\mathrm{f})\dag}G_{B}^{-1}(p)\Phi_{p}^{(\mathrm{f})}
  +J_{p}^{\dag}\Phi_{p}^{(\mathrm{f})}+\Phi_{p}^{(\mathrm{f})\dag}J_{p}
  \right) \right] \nonumber \\
  = &
  \mathcal{N} \mathrm{Det}(G_{B})
  \exp \left[
  -\frac{1}{2} \sum_{p} J_{p}^{\dag}G_{B}(p)J_{p}
  \right], \label{eq:FModGenFun}
\end{align}
where $p$ is restricted within $\Lambda'<|\bp|\leqslant\Lambda$,
and $\mathcal{N}$ absorbs irrelevant coefficients, and
$J_{p} = (\cJ_{p}, \acJ_{-p})^{T}$. The Green's functions are given by
\begin{subequations}
\begin{align}
  & G_{B}(p) \equiv
    \begin{pmatrix}
      G_{11}(p) & G_{12}(p) \\
      G_{21}(p) & G_{22}(p)
    \end{pmatrix}
    = \frac{1}{(i\omega_{n})^{2}-\omega_{B}^{2}(\bp)}
    \begin{pmatrix}
      i\omega_{n}+\frac{\bp^{2}}{2m_{B}}+g_{B}n_{B} & -g_{B}n_{B} \\
      -g_{B}n_{B} & -i\omega_{n}+\frac{\bp^{2}}{2m_{B}}+g_{B}n_{B}
    \end{pmatrix}, \\
  & \omega_{B}(\bp)
    = \sqrt{\frac{\bp^{2}}{2m_{B}}
    \left(\frac{\bp^{2}}{2m_{B}}+2g_{B}n_{B} \right)}.
\end{align}  
\end{subequations}
Note that $G_{11}(p) = G_{22}(-p)$ and
$G_{12}(p) = G_{21}(p) = G_{12}(-p) = G_{21}(-p)$.
By virtue of the generating function Eq.~(\ref{eq:FModGenFun}),
one can obtain the following Feynman rules:
\begin{subequations}
\begin{align}
  & \DefTwoPtGreenFun{(v2)}{normal GF}{(v1)}
    {edge label'=\(p_{1}{,}p_{2}\)} = 
    \langle \DefLegShort{(v)}{ff}{right}{(v0)}{edge label'=\(p_{1}\)}
    \hspace{-1.3em} 
    \DefLegShort{(v0)}{ff}{left}{(v)}{edge label'=\(p_{2}\)}
    \rangle_{\f} =
    \frac{1}{\mathcal{Z}_{\f}}
    \left. \frac{\delta^{2}\mathcal{Z}_{\f}}
    {\delta\acJ_{p_{1}}\delta\cJ_{p_{2}}}
    \right|_{\cJ=0} \hspace{-1em}=
    -G_{11}(p_{1})\delta_{p_{1}-p_{2}}, \\
  & \DefTwoPtGreenFun{(v1)}{anomalous GFout}{(v2)}
    {edge label=\(p_{1}{,}p_{2}\)} = 
    \langle \DefLegShort{(v)}{ff}{right}{(v0)}{edge label'=\(p_{1}\)}
    \hspace{-1.3em} 
    \DefLegShort{(v)}{ff}{left}{(v0)}{edge label=\(p_{2}\)}
    \rangle_{\f} =
    \frac{1}{\mathcal{Z}_{\f}}
    \left. \frac{\delta^{2}\mathcal{Z}_{\f}}
    {\delta\acJ_{p_{1}}\delta\acJ_{p_{2}}}
    \right|_{\cJ=0} \hspace{-1em}=
    -G_{12}(p_{1})\delta_{p_{1}+p_{2}}, \\
  & \DefTwoPtGreenFun{(v1)}{anomalous GFin}{(v2)}
    {edge label=\(p_{1}{,}p_{2}\)} = 
    \langle \DefLegShort{(v0)}{ff}{right}{(v)}{edge label=\(p_{1}\)}
    \hspace{-1.3em} 
    \DefLegShort{(v0)}{ff}{left}{(v)}{edge label'=\(p_{2}\)}
    \rangle_{\f} =
    \frac{1}{\mathcal{Z}_{\f}}
    \left. \frac{\delta^{2}\mathcal{Z}_{\f}}
    {\delta\cJ_{p_{1}}\delta\cJ_{p_{2}}}
    \right|_{\cJ=0} \hspace{-1em}=
    -G_{12}(p_{1})\delta_{p_{1}+p_{2}}.
\end{align}  
\end{subequations}

As a result of
$\langle\phif_{p}\rangle_{\f}  = 
\frac{1}{\mathcal{Z}_{\f}}
\left. \frac{\delta \mathcal{Z}_{\f}}{\delta\acJ_{p}}
\right|_{\cJ=0} \hspace{-1em} = 0$,
one can conclude that
\begin{align}
  \langle S_{I} \rangle_{\f} =
  \sum_{p_{1},p_{2}} \langle \cL_{I}(p_{1},p_{2}) \rangle_{\f}
  = \sum_{p_{1},p_{2}} \FDZeroLegOneLoop,
\end{align}
which is irrelevant to the slow modes.
As for the quadratic terms, we have
\begin{align}
  \langle S_{I}^{2} \rangle_{\f} - \langle S_{I} \rangle_{\f}^{2}
  & = \sum_{p_{1}\cdots p_{4}} \left[
    \langle \cL_{I}(p_{1},p_{2})\cL_{I}(p_{3},p_{4}) \rangle_{\f}
    - \langle \cL_{I}(p_{1},p_{2}) \rangle_{\f}
    \langle \cL_{I}(p_{3},p_{4}) \rangle_{\f}
    \right] \nonumber \\
  & = \sum_{p_{1}\cdots p_{4}} \left(
    2 \times \SsqContribOne{(r)}{fs,edge label'=\(p_{4}\)}{(v2)}
    {normal GF,edge label'=\(p_{2}{,}p_{3}\)}
    {(v1)}{fs,edge label'=\(p_{1}\)}{(l)}
    + \SsqContribOne{(v1)}{fs,edge label'=\(p_{1}\)}{(l)}
    {anomalous GFout,edge label'=\(p_{2}{,}p_{3}\)}
    {(v2)}{fs,edge label=\(p_{4}\)}{(r)}
    + \SsqContribOne{(l)}{fs,edge label=\(p_{1}\)}{(v1)}
    {anomalous GFin,edge label'=\(p_{2}{,}p_{3}\)}
    {(r)}{fs,edge label'=\(p_{4}\)}{(v2)}
    \right) + C, \label{eq:SM_SsqContribOne}
\end{align}
where $C$ is a constant that used to absorb all slow-mode-irrelevant
terms.

\section{Renormalization Group Flow Equations}

From the macroscopic point of view, the BEC with two impurities has an
induced energy $\omega_{I}$,
and the condensate density $n_{B}$ is not altered by the presence of
the impurities.
Here, we also neglect the self-localization of the impurities
and deformations of the BEC as Ref.~\cite{PhysRevA.88.053632}.
We make the simplifying assumption that
$\aphi(0,\omega)\phi(0,\omega) =
\aphi(0,\omega)\aphi(0,-\omega) =
\phi(0,-\omega)\phi(0,\omega) =
n_{B}\delta(\omega-\omega_{I})$.

The terms in Eq.~(\ref{eq:SM_SsqContribOne}) that align with the
original action Eq.~(\ref{eq:SM_S_Lambda}) are extracted and divided
into two parts, $T_{g}$ and $T_{\Sigma}$, as
\begin{align}
  & \langle S_{I}^{2} \rangle_{\f} - \langle S_{I} \rangle_{\f}^{2}
    \approx C + T_{g} + T_{\Sigma}, \\
  & T_{g} = 
    -\frac{2}{V} \hspace{-0.8em}
    \sum_{\omega_{n},\bp_{1},\bp_{2}} \hspace{-0.5em}
    (1-\delta_{\bp_{1}}\delta_{\bp_{2}})    
    \left[ \frac{g_{I}^{2}}{V} \hspace{-0.8em}
    \sum_{\Lambda'<|\bq|\leqslant\Lambda}\hspace{-0.8em}
    U(\bq-\bp_{1})U(\bp_{2}-\bq)G_{11}(\bq,i\omega_{n}) 
    \right]\aphis_{p_{1}}\phis_{p_{2}} , \label{eq:SM_Def_T_g} \\
  & T_{\Sigma} =
    -\beta\frac{2g_{I}^{2}n_{B}}{V}
    \hspace{-0.5em}\sum_{\Lambda'<|\bq|\leqslant\Lambda}\hspace{-0.5em}
    U(\bq)U(-\bq) G(\bq,\omega_{I}), \label{eq:SM_Def_T_Sigma}
\end{align}  
where we have defined
\begin{align}
  G(\bq,\omega_{I}) =
  G_{11}(\bq,\omega_{I}) + G_{12}(\bq,\omega_{I}). \label{eq:SM_Def_G}
\end{align}
Here $T_{g}$ includes the first term in
Eq.~(\ref{eq:SM_SsqContribOne}) with nonzero slow modes,
and $T_{\Sigma}$ comes from the three leading terms in
Eq.~(\ref{eq:SM_SsqContribOne}) with slow modes being zero.

Notice that
\begin{align}
  & \sum_{\bq} U(\bq-\bp_{1})U(\bp_{2}-\bq)G_{11}(\bq,i\omega_{n})
    = \left[\sum_{\bq}\left(
    1+e^{i\bq\cdot\bR}\frac{\cos[(\bp_{2}+\bp_{1})\cdot\bR/2]}
    {\cos[(\bp_{2}-\bp_{1})\cdot\bR/2]}
    \right)G_{11}(\bq,i\omega_{n})\right] U(\bp_{2}-\bp_{1}),
    \label{eq:SM_U_Simplify_1} \\
  & \sum_{\bq}U(\bq)U(-\bq) G(\bq,\omega_{I})
    = 2\sum_{\bq}\left[1+\cos(\bq\cdot\bR)\right] G(\bq,\omega_{I})
    = 2\sum_{\bq}\left(1+e^{i\bq\cdot\bR}\right) G(\bq,\omega_{I}),
    \label{eq:SM_U_Simplify_2}
\end{align}
where $\bR \equiv \br_{1}-\br_{2}$.
Substituting Eqs.~(\ref{eq:SM_Def_T_g})-(\ref{eq:SM_U_Simplify_2})
into Eq.~(\ref{eq:SM_S_Lambda_Prime_Form}), then the effective action
$S_{\Lambda'}$ can be written as
\begin{align}
  S_{\Lambda'} = \frac{1}{2}
    \hspace{-0.8em}
    \sum_{\substack{\omega_{n} \\ 0<|\bp|\leqslant\Lambda'}}
  \hspace{-0.8em}
  \Phi_{p}^{(\s)\dag}[-G_{B}^{-1}(p)]\Phi_{p}^{(\s)}
  + \beta\Sigma_{I}(\Lambda') + \frac{g_{I}(\Lambda')}{V}
  \hspace{-1.5em}
  \sum_{\substack{\omega_{n} \\ |\bp_{1}|,|\bp_{2}|\leqslant\Lambda'}}
  \hspace{-1em}
  (1-\delta_{\bp_{1}}\delta_{\bp_{2}})    
  U(\bp_{2}-\bp_{1})\aphis_{p_{1}}\phis_{p_{2}},
\end{align}
where
\begin{align}
  & g_{I}(\Lambda') = g_{I}(\Lambda)
    + \frac{g_{I}^{2}}{V} \hspace{-0.5em}
    \sum_{\Lambda'<|\bq|\leqslant\Lambda}\hspace{-0.5em}
    \left(
    1+e^{i\bq\cdot\bR}\frac{\cos[(\bp_{2}+\bp_{1})\cdot\bR/2]}
    {\cos[(\bp_{2}-\bp_{1})\cdot\bR/2]}
    \right)G_{11}(\bq,i\omega_{n}), \label{eq:SM_g_new} \\
  & \Sigma_{I}(\Lambda') = \Sigma_{I}(\Lambda)
    + \frac{2g_{I}^{2}n_{B}}{V}
    \hspace{-0.5em}\sum_{\Lambda'<|\bq|\leqslant\Lambda}\hspace{-0.5em}
    \left(1+e^{i\bq\cdot\bR}\right)
    G(\bq,\omega_{I}). \label{eq:SM_Sigma_new} 
\end{align}
As the $\Delta\Lambda$ becoming infinitesimal,
Eqs.~(\ref{eq:SM_g_new})-(\ref{eq:SM_Sigma_new}) yield the flow
equations
\begin{align}
  &   \frac{\ud g_{I}}{\ud\Lambda} = 
    - \frac{g_{I}^{2}\Lambda^{2}}{(2\pi)^{3}}\int \ud\Omega
    \left(
    1+e^{i\Lambda R \cos\theta}\frac{\cos[(\bp_{2}+\bp_{1})\cdot\bR/2]}
    {\cos[(\bp_{2}-\bp_{1})\cdot\bR/2]}
    \right)G_{11}(\Lambda,i\omega_{n}), \label{eq:SM_dg_dLambda} \\
  & \frac{\ud\Sigma_{I}}{\ud\Lambda} =
    - \frac{2g_{I}^{2}n_{B}\Lambda^{2}}{(2\pi)^{3}}
    \int \ud\Omega
    \left(1+e^{i\Lambda R \cos\theta}\right)
    G(\Lambda,\omega_{I}), \label{eq:SM_dSigma_dLambda}
\end{align}
where $\ud\Omega = \sin\theta\ud\theta\ud\varphi$ is the differential
solid angle, $R\equiv|\bR|=|\br_{1}-\br_{2}|$ is the distance between
two impurities.

To further simplify the flow equations, we require that
$\Sigma_{I}(\Lambda) \approx 2g_{I}(\Lambda)n_{B}$, so that
Eq.~(\ref{eq:SM_Sigma_new}) reduces to
\begin{align}
  \frac{\ud\Sigma_{I}}{\ud\Lambda} 
  \approx
  -\frac{\Sigma_{I}^{2}\Lambda^{2}}{2n_{B}(2\pi)^{3}}
  \int \ud\Omega \left( 1+e^{i\Lambda R \cos\theta} \right)
  G(\Lambda,\omega_{I}). \label{eq:SM_dSigma_dLambda_Sim}
\end{align}
Solving Eq.~(\ref{eq:SM_dSigma_dLambda_Sim}) leads to
\begin{align}
  \frac{1}{\Sigma_{I}(\Lambda\rightarrow0)}
  = \frac{1}{\Sigma_{I}(\Lambda)} - \frac{1}{2n_{B}V}
  \sum_{|\bq|\leqslant\Lambda} \left( 1+e^{i\bq\cdot\bR}\right)
  G(\bq,\omega_{I}), \label{eq:SM_Sigma_Eq}
\end{align}
where $\Sigma_{I}(\Lambda\rightarrow0) = \omega_{I} =
\Sigma_{I}(\omega_{I})$, and
\begin{align}
  \frac{1}{\Sigma_{I}(\Lambda)} \approx
  \frac{1}{2g_{I}(\Lambda)n_{B}}
  = \frac{1}{2n_{B}} \left( \frac{m_{B}}{2\pi a_{I}}
  -\frac{1}{V}\sum_{|\bq|\leqslant\Lambda} \frac{2m_{B}}{\bq^{2}}
  \right).
\end{align}
Hence the self-energy $\Sigma_{I}(\omega)$ reads
\begin{align}
  \Sigma_{I}(\omega)
  =  2n_{B}\left[
  \frac{m_{B}}{2\pi a_{I}}
  - \frac{1}{V} \sum_{|\bq|\leqslant\Lambda} \left(
  \left( 1+e^{i\bq\cdot\bR}\right)
  G(\bq,\omega) + \frac{2m_{B}}{\bq^{2}}
  \right) \right]^{-1}. \label{eq:SM_Sigma_2Imp}
\end{align}

To examine this result, we consider the case of single impurity
immersed in a BEC. By applying transformations $\bR \rightarrow 0$
and $2g_{I}\rightarrow g_{I}$, the double-impurity action
Eq.~(\ref{eq:SM_S_Lambda}) will reduce to the single-impurity model.
Then we apply the transformations to Eq.~(\ref{eq:SM_Sigma_Eq}), we
then obtain the self-energy for the single impurity
\begin{align}
  \Sigma_{I}(\omega)
  =  n_{B}\left[
  \frac{m_{B}}{2\pi a_{I}}
  - \frac{1}{V} \sum_{|\bq|\leqslant\Lambda} \left(
  G(\bq,\omega) + \frac{2m_{B}}{\bq^{2}}
  \right) \right]^{-1}. \label{eq:SM_Sigma_1Imp}
\end{align}
Comparing with the self-energy in Ref.~\cite{PhysRevA.88.053632},
we neglect the dressing of the impurity with depleted bosons, that is,
the second term of Eq.~(7) in reference.
However, we additionally consider the anomalous propagator in
Eq.~(\ref{eq:SM_Sigma_1Imp}) since $G(q)=G_{11}(q)+G_{12}(q)$
(compared to Eq.~(9) in reference).

\section{Complex Analysis of Self-Energy
  $\Sigma_{I}(\omega)$}
In this section, we introduce the healing length
$\xi=1/(\lambda n_{B}^{1/3})$ with the dimensionless quantity
$\lambda$ defined as $\lambda \equiv \sqrt{8\pi a_{B}n_{B}^{1/3}}$.
For simplicity, we denote $R/\xi \rightarrow R$,
$q\xi \rightarrow q$, $\omega/(g_{B}n_{B}) \rightarrow \omega$ and
$\omega_{B}(q)/(g_{B}n_{B}) \rightarrow \omega_{B}(q)$.
Then the self-energy for double impurities
Eq.~(\ref{eq:SM_Sigma_2Imp}) becomes
\begin{align}
  \Sigma_{I}(\omega,R) =
  \left(\frac{\hbar^{2}n_{B}^{2/3}}{m_{B}} \right)
  \left( \frac{1}{4\pi a_{I}n_{B}^{1/3}}
  + \frac{\lambda}{2\pi^{2}} \left[
  L(\omega) + I(\omega,R)
  \right]\right)^{-1}, \label{eq:SM_Sigma_NoDim}
\end{align}
where we have defined dimensionless functions
\begin{align}
  & L(\omega) =
    \int_{0}^{\infty} dq
    \frac{(\omega-2)q^{2}+\omega^{2}}{q^{2}(q^{2}+2)-\omega^{2}},
  \label{eq:SM_Def_L} \\
  & I(\omega,R) =
    \int_{0}^{\infty} dq~q^{2}
    \frac{\sin(qR)}{qR}
    \frac{\omega+q^{2}}{q^{2}(q^{2}+2)-\omega^{2}}.
    \label{eq:SM_Def_I}
\end{align}
Here we have taken $\Lambda \rightarrow \infty$.
We want to extract the real and imaginary pars of
$\Sigma_{I}(\omega+i0^{+})$.
To this end, we now investigate $L(\omega+i0^{+})$ and
$I(\omega+i0^{+},R)$. We denote the real parts as $\rRe L(\omega)$ and
$\rRe I(\omega,R)$, and imaginary parts as $\rIm L(\omega)$ and
$\rIm I(\omega,R)$.
By taking advantage of
\begin{align}
  \frac{1}{x \pm i0^{+}} = \mathcal{P}\left(\frac{1}{x}\right)
  \mp i\pi\delta(x),
\end{align}
one obtains
\begin{align}
  & \rRe L(\omega) =
    \mathcal{P}\int_{0}^{\infty} dq
    \frac{(\omega-2)q^{2}+\omega^{2}}{q^{2}(q^{2}+2)-\omega^{2}}, \\
  & \rRe I(\omega,R) =
    \mathcal{P}\int_{0}^{\infty} dq~q^{2}
    \frac{\sin(qR)}{qR}
    \frac{\omega+q^{2}}{q^{2}(q^{2}+2)-\omega^{2}}, \\
  & \rIm L(\omega) =
    \pi\sgn(\omega)\int_{0}^{\infty} dq
    \left[(\omega-2)q^{2}+\omega^{2}\right]
    \delta\left[q^{2}(q^{2}+2)-\omega^{2}\right], \\
  & \rIm I(\omega,R) =
    \pi\sgn(\omega)\int_{0}^{\infty} dq~q^{2}
    \frac{\sin(qR)}{qR}
    (\omega+q^{2}) \delta\left[q^{2}(q^{2}+2)-\omega^{2}\right],
\end{align}
where $\sgn(x)$ is the sign function.
Note that equation $q^{2}(q^{2}+2)-\omega^{2} = 0$ has four solutions:
$\pm q_{-}$ and $\pm iq_{+}$, where
\begin{align}
  q_{\pm} \equiv \sqrt{\sqrt{\omega^{2}+1} \pm 1} >0,
\end{align}
provided that $\omega \neq 0$. Then the imaginary parts can be expressed as
\begin{align}
  & \rIm L(\omega) =
    \sgn(\omega) \frac{\pi}{4}
    \frac{(\omega-2)q_{-}^{2}+\omega^{2}}{q_{-}\sqrt{\omega^{2}+1}},
    \label{eq:SM_ImL} \\ 
  & \rIm I(\omega,R) =
    \sgn(\omega) \frac{\pi}{4}
    \left(1+\frac{\omega-1}{\sqrt{\omega^{2}+1}}\right)
    \frac{\sin(q_{-}R)}{R}. \label{eq:SM_ImI}
\end{align}

Since the integrands are even functions, the real parts can be also
written as
\begin{align}
  & \rRe L(\omega) =
    \mathcal{P} \int_{-\infty}^{\infty} dq
    \frac{(\omega-2)q^{2}+\omega^{2}}{2[q^{2}(q^{2}+2)-\omega^{2}]}, \\
  & \rRe I(\omega,R) =
    \mathcal{P} \int_{-\infty}^{\infty} dq~q^{2}
    \frac{\sin(qR)}{qR}
    \frac{\omega+q^{2}}{2[q^{2}(q^{2}+2)-\omega^{2}]}.
\end{align}

\begin{figure}[!htbp]
\centering
\begin{tikzpicture}[scale=0.35,line width=0.8pt,
  bent arrow at/.style = {decoration={markings,
      mark=at position {#1*\pgfdecoratedpathlength + 2} with
      {\coordinate (bent arrow 1);},
      mark=at position {#1*\pgfdecoratedpathlength + 3} with
      {\coordinate (bent arrow 2);},
      mark=at position {#1*\pgfdecoratedpathlength + 4} with
      {\coordinate (bent arrow 3);},
      mark=at position {#1*\pgfdecoratedpathlength + 5} with
      {\coordinate (bent arrow 4);
        \draw[-{Latex[length=7pt,bend]}] (bent arrow 1)
        to[curve through={(bent arrow 2) .. (bent arrow 3)}]
        (bent arrow 4);}
    }}
  ]
  \draw[-{Stealth}] (-10,0)--(10,0) node[right] {Re$(z)$};
  \draw[-{Stealth}] (0,-5)--(0,10) node[above] {Im$(z)$};
  \draw[fill] (0,4) circle[radius=0.2] (1,4) node[right] {$iq_{+}$};
  \draw[fill] (0,-4) circle[radius=0.2] node[right] {$-iq_{+}$};
  \draw[fill] (-4,0) circle[radius=0.2] node[below] {$-q_{-}$};
  \draw[fill] (4,0) circle[radius=0.2] (4.2,-0.2) node[below] {$q_{-}$};  
  \draw[postaction={bent arrow at=0.5,decorate},blue]
  (9,0) to [out=90,in=0] (0,9);
  \draw[postaction={bent arrow at=0.5,decorate},blue]
  (0,9) to [out=180,in=90] (-9,0);
  \draw[postaction={bent arrow at=0.5,decorate},blue] (-9,0) -- (-5,0);
  \draw[postaction={bent arrow at=0.5,decorate},blue]
  (-5,0) arc [radius=1, start angle=180, end angle=0];
  \draw[postaction={bent arrow at=0.5,decorate},blue] (-3,0) -- (0,0);
  \draw[postaction={bent arrow at=0.5,decorate},blue] (0,0) -- (3,0);
  \draw[postaction={bent arrow at=0.5,decorate},blue]
  (3,0) arc [radius=1, start angle=180, end angle=0];
  \draw[postaction={bent arrow at=0.5,decorate},blue] (5,0) -- (9,0);
  \draw[postaction={bent arrow at/.list={0.25,0.75},decorate},blue]
  (0,4) circle [radius=1];
  \draw (7,7) node[right,blue] {$C$};
  \draw (-4,1) node[above,blue] {$\varepsilon_{1}$};
  \draw (4,1) node[above,blue] {$\varepsilon_{2}$};
  \draw (-1,4) node[left,blue] {$\gamma$};  
\end{tikzpicture}
\caption{The integral paths. The radius of path $C$ is extended to
  infinity, $R_{C}\rightarrow\infty$.
  The radii of path $\varepsilon_{1}$, $\varepsilon_{2}$ and $\gamma$
  are approaching to zero,
  $R_{\varepsilon_{1}},R_{\varepsilon_{2}},R_{\gamma} \rightarrow 0$.}
\label{fig:SM_IntegralPath}
\end{figure}
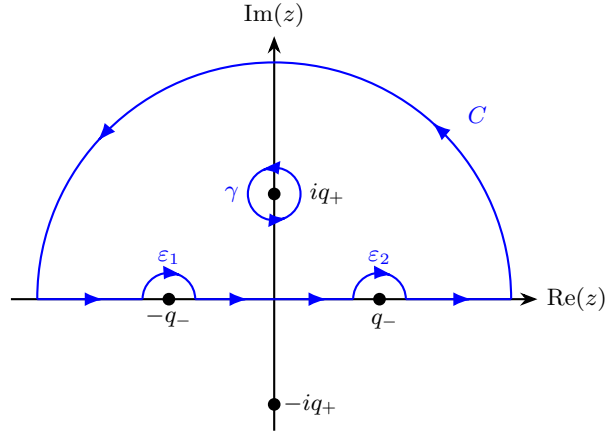

Note that the integrands are analytic on the complex plane except for
the four poles, $\pm q_{-}$ and $\pm iq_{+}$, as shown in
FIG.~\ref{fig:SM_IntegralPath}.
We construct contour integrals as
\begin{align}
  & \left( \mathcal{P} \int_{-\infty}^{\infty} + \int_{C} +
    \int_{\varepsilon_{1}} + \int_{\varepsilon_{2}} \right)
    dz \frac{(\omega-2)z^{2}+\omega^{2}}{2[z^{2}(z^{2}+2)-\omega^{2}]}
    = \oint_{\gamma} dz
    \frac{(\omega-2)z^{2}+\omega^{2}}{2[z^{2}(z^{2}+2)-\omega^{2}]}, \\ 
  & \left( \mathcal{P} \int_{-\infty}^{\infty} + \int_{C} +
    \int_{\varepsilon_{1}} + \int_{\varepsilon_{2}} \right)
    dz~z^{2} \frac{\sin(zR)}{zR}
    \frac{\omega+z^{2}}{2[z^{2}(z^{2}+2)-\omega^{2}]}
    = \oint_{\gamma} dz~z^{2} \frac{\sin(zR)}{zR}
    \frac{\omega+z^{2}}{2[z^{2}(z^{2}+2)-\omega^{2}]},
\end{align}
where the integral paths are marked in FIG.~\ref{fig:SM_IntegralPath}.
Above contour integrals lead to
\begin{align}
  & \rRe L(\omega) =
    \left( \oint_{\gamma} - \int_{C} - \int_{\varepsilon_{1}} -
    \int_{\varepsilon_{2}} \right) 
    dz\frac{(\omega-2)z^{2}+\omega^{2}}{2[z^{2}(z^{2}+2)-\omega^{2}]},
  \\  
  & \rRe I(\omega,R) =
    \left( \oint_{\gamma} - \int_{C} - \int_{\varepsilon_{1}} -
    \int_{\varepsilon_{2}} \right) 
    dz~z^{2} \frac{\sin(zR)}{zR}
    \frac{\omega+z^{2}}{2[z^{2}(z^{2}+2)-\omega^{2}]}.
\end{align}
As the result of the residue theorem, we finally obtain
\begin{align}
  & \rRe L(\omega) =
    \frac{\pi}{4}
    \frac{(\omega-2)q_{+}^{2}-\omega^{2}}{q_{+}\sqrt{\omega^{2}+1}}, 
    \label{eq:SM_ReL} \\
  & \rRe I(\omega,R) =
    \frac{\pi}{4} \left[
    \left(1-\frac{\omega-1}{\sqrt{\omega^{2}+1}}\right)
    \frac{e^{-q_{+}R}}{R} +
    \left(1+\frac{\omega-1}{\sqrt{\omega^{2}+1}}\right)
    \frac{\cos(q_{-}R)}{R} 
    \right].
    \label{eq:SM_ReI}
\end{align}
Bring real parts Eqs.~(\ref{eq:SM_ReL})-(\ref{eq:SM_ReI}) and
imaginary parts Eqs.~(\ref{eq:SM_ImL})-(\ref{eq:SM_ImI}) all together
leads to
\begin{align}
  & L(\omega+i0^{+}) =
    \frac{\pi}{4} \left[
    \frac{(\omega-2)q_{+}^{2}-\omega^{2}}{q_{+}\sqrt{\omega^{2}+1}}
    +i\sgn(\omega)
    \frac{(\omega-2)q_{-}^{2}+\omega^{2}}{q_{-}\sqrt{\omega^{2}+1}}
    \right], \\
  & I(\omega+i0^{+},R) =
    \frac{\pi}{4} \left[
    \left(1-\frac{\omega-1}{\sqrt{\omega^{2}+1}}\right)
    \frac{e^{-q_{+}R}}{R} +
    \left(1+\frac{\omega-1}{\sqrt{\omega^{2}+1}}\right)
    \frac{e^{i\sgn(\omega)q_{-}R}}{R} 
    \right].
\end{align}

\section{Derivation of Effective Potentials}

In this section, we put all the units back to the self-energy, leading
to
\begin{align}
  \Sigma_{I}(\omega,R)
  &= \left(\frac{\hbar^{2}n_{B}^{2/3}}{m_{B}} \right)
  \left( \frac{1}{4\pi a_{I}n_{B}^{1/3}}
  + \frac{\lambda}{2\pi^{2}} \left[
  L(\omega) + I(\omega,R)
  \right]\right)^{-1} \nonumber \\
  &= \left(\frac{4\pi\hbar^{2}n_{B}}{m_{B}} \right)
  \left( \frac{1}{a_{I}}
  + \frac{2}{\pi} \sqrt{\frac{2m_{B}}{\hbar^{2}}}
  [\sqrt{g_{B}n_{B}}L(\omega)] +
  \frac{2}{\pi} [\xi^{-1}I(\omega,R)] \right)^{-1}, 
\end{align}
where $L(\omega +i0^{+})$ and $I(\omega+i0^{+},R)$ are still
dimensionless functions, now expressed as
\begin{align}
  & L(\omega +i0^{+}) =
    \frac{\pi}{4\sqrt{g_{B}n_{B}}} \left[
    \frac{(\omega-2g_{B}n_{B})\frac{\hbar^{2}q_{+}^{2}}{2m_{B}}-\omega^{2}}
    {\frac{\hbar q_{+}}{\sqrt{2m_{B}}}\sqrt{\omega^{2}+(g_{B}n_{B})^{2}}}
    +i\sgn(\omega)
    \frac{(\omega-2g_{B}n_{B})\frac{\hbar^{2}q_{-}^{2}}{2m_{B}}+\omega^{2}}
    {\frac{\hbar q_{-}}{\sqrt{2m_{B}}}\sqrt{\omega^{2}+(g_{B}n_{B})^{2}}}
    \right], \\
  & I(\omega+i0^{+},R) =
    \frac{\pi}{4} \left[
    \left(1-\frac{\omega-g_{B}n_{B}}
    {\sqrt{\omega^{2}+(g_{B}n_{B})^{2}}}\right)
    \frac{e^{-q_{+}R}}{R/\xi} +
    \left(1+\frac{\omega-g_{B}n_{B}}
    {\sqrt{\omega^{2}+(g_{B}n_{B})^{2}}}\right)
    \frac{e^{i\sgn(\omega)q_{-}R}}{R/\xi} 
    \right], \\
  & q_{\pm} =
    \sqrt{\frac{2m_{B}}{\hbar^{2}}
    \left[\sqrt{\omega^{2}+(g_{B}n_{B})^{2}} \pm g_{B}n_{B}\right]}.
\end{align}

In the case of ideal BEC, the boson-boson interaction vanishes as
$g_{B}n_{B} = 0$, leading to
$ q_{\pm} = \sqrt{2m_{B}|\omega|/\hbar^{2}}$, and
\begin{align}
  \Sigma_{I}(\omega,R) =
  \left(\frac{4\pi\hbar^{2}n_{B}}{m_{B}} \right)
  \left[ \frac{1}{a_{I}} +
  \left(\frac{e^{i\kappa(\omega)R}}{R} + i\kappa(\omega)\right)
  \Theta(\omega) +
  \left(\frac{e^{-\kappa(\omega)R}}{R} - \kappa(\omega)\right)
  \Theta(-\omega) \right]^{-1}, 
\end{align}
where we have defined
$\kappa(\omega) \equiv \sqrt{2m_{B}|\omega|/\hbar^{2}} > 0$,
and $\Theta(x)$ is the Heaviside step function.
Note that $\Sigma_{I}(\omega,R)$ has real solutions when $\omega<0$,
which indicates a bound state for the ideal BEC.
To prevent the collapse of BEC, the ground-state energy should be
finite and small, so that the energy equation reduced to
$\omega_{I} = \rRe\Sigma_{I}(0)$, leading to~\cite{Panochko_2021}
\begin{align}
  \omega_{BEC} = \frac{4\pi\hbar^{2}n_{B}}{m_{B}} \frac{1}{a_{I}^{-1} + R^{-1}},
\end{align}
which requires $a_{I}<0$ and $R>|a_{I}|$. This energy generates
so-called ``shifted Newtonian'' attractive
potential~\cite{PhysRevA.107.063301}
\begin{align}
  V_{SN}(R) = -\frac{4\pi\hbar^{2}n_{B}a_{I}^{2}}{m_{B}}\frac{1}{a_{I} + R}.
\end{align}

If the ideal BEC collapses, then $\omega_{I} \rightarrow -\infty$,
which leads to $[\rRe\Sigma_{I}(\omega_{I})]^{-1}=0$, and results in
equation
\begin{align}
  \frac{1}{a_{I}} + \frac{e^{-\kappa(\omega)R}}{R} - \kappa(\omega)
  = 0. \label{eq:SM_Eq_3BodyEnergy}
\end{align}
Solving Eq.~(\ref{eq:SM_Eq_3BodyEnergy}), we obtain
\begin{align}
  \kappa(\omega)
   = \frac{1}{a_{I}} + \frac{1}{R}W\left(e^{-R/a_{I}}\right),
\end{align}
then
\begin{equation}
  \omega_{b} = -\frac{\hbar^{2}}{2m_{B}} \left[
    \frac{1}{a_{I}} + \frac{1}{R}
    W\left(e^{-R/a_{I}}\right)
  \right]^{2}, \label{eq:SM_Three-Body_Binding_Energy}
\end{equation}
where $a_{I}^{-1}>-R^{-1}$ and $W(x)$ is the Lambert $W$
function~\cite{PhysRevA.79.013629, PhysRevA.102.063321,
  PhysRevA.107.063301}.
Expanding $\omega_{b}$ with respect to $R$ around $0$ provides
\begin{align}
  \omega_{b} \approx -\frac{\hbar^{2}}{2m_{B}}
  \left[ -\frac{W^{2}(1)}{R^{2}}
  - \frac{2W(1)}{a_{I}[1+W(1)]R}
  - \frac{1+W(1)+W^{2}(1)}{a_{I}^{2}[1+W(1)]^{3}} + O(R)
  \right].
\end{align}
Hence in the large scattering length limit and at short distance,
this energy yields the Efimov
attraction~\cite{doi:10.7566/JPSJ.87.043002},
\begin{align}
V_{E}(R) = -\frac{\hbar^{2}}{2m_{B}}\frac{W^{2}(1)}{R^{2}}. \label{eq:SM_V_Efimov}
\end{align}

In the case of the weakly interacting BEC, we consider
$|\omega| \gg g_{B}n_{B}$ and $\omega<0$, then
$q_{+} \approx q_{-} \approx \kappa(\omega)$,
and the self-energy becomes
\begin{align}
  \Sigma_{I}(\omega,R)
  &\approx \left(\frac{\hbar^{2}n_{B}^{2/3}}{m_{B}} \right)
    \left[ \frac{1}{4\pi a_{I}n_{B}^{1/3}}
    + \frac{\lambda}{4\pi} \left(
    -\sqrt{\frac{|\omega|}{g_{B}n_{B}}} +
    \frac{e^{-\kappa(\omega)R}}{R/\xi} \right)\right]^{-1}
    = \frac{4\pi\hbar^{2}n_{B}}{m_{B}} \left(
    \frac{1}{a_{I}} - \kappa(\omega) + \frac{e^{\kappa(\omega)R}}{R}
    \right)^{-1},
\end{align}
which will also lead to the three-body binding energy
Eq.~(\ref{eq:SM_Three-Body_Binding_Energy}) and the Efimov attraction
Eq.~(\ref{eq:SM_V_Efimov}).

In the case of $|\omega| \ll g_{B}n_{B}$, we have
$q_{+} \approx \sqrt{4m_{B}g_{B}n_{B}/\hbar^{2}} = \sqrt{2}/\xi$
and $q_{-} \approx 0$,
and the self-energy can be approximated as
\begin{align}
  \Sigma_{I}(\omega,R)
  &\approx \left(\frac{\hbar^{2}n_{B}^{2/3}}{m_{B}} \right)
    \left[ \frac{1}{4\pi a_{I}n_{B}^{1/3}}
    + \frac{\lambda}{4\pi} \left(
    -\sqrt{2} + \frac{e^{-\sqrt{2}R/\xi}}{R/\xi} \right)\right]^{-1}, 
\end{align}
so that impurity-induced energy reads~\cite{atoms10010019}
\begin{equation}
  \omega_{\mathrm{Weak}} =
  \frac{4\pi\hbar^{2}n_{B}^{2/3}}{m_{B}}
  \left[ \frac{1}{a_{I}n_{B}^{1/3}} - \lambda\left(
      \sqrt{2}-\frac{e^{-\sqrt{2}R/\xi}}{R/\xi}
    \right)\right]^{-1}.
  \label{eq:SM_Weak_Int_Energy}
\end{equation}
For the attractive polaron ($\omega_{\mathrm{Weak}}<0$),
$a_{I}n_{B}^{1/3} <0$, if $R/\xi \gg 1$ and $|a_{I}n_{B}^{1/3}|\ll1$,
then Eq.~(\ref{eq:SM_Weak_Int_Energy}) can be approximated as
\begin{align}
  \omega_{\mathrm{Weak}} =
  \frac{4\pi\hbar^{2}n_{B}a_{I}}{m_{B}}
  \sum_{n=0}^{\infty} \left[ (\lambda a_{I}n_{B}^{1/3})
  \left(\sqrt{2} - \frac{e^{-\sqrt{2}R/\xi}}{R/\xi}
  \right)\right]^{n}
  \approx \frac{4\pi\hbar^{2}n_{B}a_{I}}{m_{B}}
  \left[ 1+ \lambda a_{I}n_{B}^{1/3}
  \left(\sqrt{2} - \frac{e^{-\sqrt{2}R/\xi}}{R/\xi}
  \right)\right],
\end{align}
which results in a well-known Yukawa
potential~\cite{PhysRevX.8.031042, doi:10.7566/JPSJ.87.043002,
  Jager_2022, PhysRevA.107.063301}
\begin{align}
  V_{Y} =
  -\frac{4\pi\hbar^{2}n_{B}a_{I}}{m_{B}}  
  \frac{e^{-\sqrt{2}R/\xi}}{R/\xi}.
\end{align}
And if $R/\xi \ll 1$, Eq.~(\ref{eq:SM_Weak_Int_Energy}) can be
approximated as
\begin{equation}
  \omega_{\mathrm{Weak}} \approx
  \frac{4\pi\hbar^{2}n_{B}R}{m_{B}} + O(R^{2}),\label{eq:SM_E_rep}
\end{equation}
which corresponds to the repulsive polaron
($\omega_{\mathrm{Weak}}>0$), and is independent of the sign of
$a_{I}n_{B}^{1/3}$.
Eq.~(\ref{eq:SM_E_rep}) also shows that impurity-induced energy
$\omega_{\mathrm{Weak}}$ linearly approaches zero as $R \rightarrow
0$, so that two impurities are asymptotically free.

\end{document}